\documentclass[aps,prd,superscriptaddress,nofootinbib,fixlfloat, 12pt]{revtex4-1}
\usepackage{graphicx}
\usepackage{dcolumn}
\usepackage{bm}
\usepackage{natbib}

\newcommand{\be}{\begin{equation}}
\newcommand{\ee}{\end{equation}}
\newcommand{\bea}{\begin{eqnarray}}
\newcommand{\eea}{\end{eqnarray}}

\newcommand{\Fermi}{{\slshape Fermi}}

\newcommand{\degree}{^{\rm o}}

\newcommand{\zrock}{$Z_{\rm rock}$}

\begin{document}

\title{Closing in on the Fermi Line with a New Observation Strategy}

\author{Christoph Weniger}
\affiliation{GRAPPA Institute, Univ.~of Amsterdam, Science Park 904, 1098 GL
Amsterdam, Netherlands}

\author{Meng Su}
\affiliation{Department of Physics, and Kavli Institute for Astrophysics and
  Space Research, Massachusetts Institute of Technology, Cambridge, MA 02139,
  USA}
\affiliation{Institute for Theory and Computation,
  Harvard-Smithsonian Center for Astrophysics, 
  60 Garden Street, MS-51, Cambridge, MA 02138, USA} 
\affiliation{Einstein Fellow}

\author{Douglas P.~Finkbeiner}
\affiliation{Institute for Theory and Computation,
  Harvard-Smithsonian Center for Astrophysics, 
  60 Garden Street, MS-51, Cambridge, MA 02138, USA} 
\affiliation{Center for the Fundamental Laws of Nature,
  Physics Department, 
  Harvard University, 
  Cambridge, MA 02138, USA}

\author{Torsten Bringmann}
\affiliation{II.~Institute for Theoretical Physics, University of Hamburg,
Luruper Chaussee 149, 22761 Hamburg, Germany}

\author{Nestor Mirabal}
\affiliation{Universidad Complutense de Madrid, Spain}

\begin{abstract}
  Evidence for a spectral line in the inner Galaxy has caused a great deal of
  excitement over the last year, mainly because of its interpretation as a
  possible dark matter signal.  The observation has raised important questions
  about statistics and suspicions about systematics, especially in photons
  from the Earth limb.  With enough additional data, we can address these
  concerns.  In this white paper, we summarize the current
  observational situation and project future sensitivities, finding that the
  status quo is dangerously close to leaving the issue unresolved until 2015.  We
  advocate a change in survey strategy that more than doubles the data rate in
  the inner Galaxy, and is relatively non-disruptive to other survey science.  
  This strategy will clearly separate the null hypothesis from the line signal
  hypothesis and provide ample limb data for systematics checks by the end of
  2014. The standard survey mode may not. 
\end{abstract}


\maketitle

\section{Introduction}
For indirect dark matter searches with gamma rays, discriminating between a
signal from conventional astrophysical backgrounds is challenging (for a
recent review see~\cite{Bringmann:2012ez}).  Among various possible
signatures, gamma-ray line emission is a long-sought ``smoking gun'' for dark
matter annihilation~\cite{Bergstrom:1988fp}, as no plausible astrophysical
background can produce a diffuse line signature.
The first claims for a spectral feature around 130 GeV in the LAT data were made by Bringmann
\textit{et al.}~\citep{Bringmann:2012} and Weniger~\citep{Weniger:2012}, who
performed a spectral fit to photon events in regions of interest in the inner
Galaxy designed to maximize S/N. They found a line structure at 130 GeV with
4.6$\sigma$ significance, or 3.2$\sigma$ after the trials factor
correction~\citep{Weniger:2012}  (for previous studies
see~\cite{Pullen:2006sy, Abdo:2010nc, Vertongen:2011mu, Ackermann:2012qk}).
This claim was quickly followed up and disputed by a number of
groups~\cite{tempel:2012ey, Boyarsky:2012ca}.  Subsequent work by Su \&
Finkbeiner~\citep{linepaper} approached the problem with template fitting, which takes into
account the spatial distribution of events along with spectral information,
and found 6.6$\sigma$ (5.1$\sigma$ after the trials factor correction) for an
Einasto profile centered $1.5\degree$ west of the Galactic center, and found
that there may be two lines, at about 111 and 129
GeV (as earlier pointed out in Ref.~\cite{Rajaraman:2012}). The lower
energy line is tantalizing because it matches the expected energy of a
$Z\gamma$ line if the higher energy is the $\gamma\gamma$ line.  These
findings have inspired a number of models and further analysis of the \Fermi\
data~\citep{Dudas:2012, Choi:2012,
Kyae:2012, Lee:2012, Rajaraman:2012, Acharya:2012, Ibarra:2012,
Buckley:2012, Chu:2012, Kang:2012, Buchmuller:2012, Bergstrom:2012b,
Heo:2012, Park:2012, Tulin:2012, Asano:2012zv, Cline:2012, Weiner:2012,
WeinerYavin:2012b,
FanReece:2012, Huang:2012, Whiteson:2012, Buchmuller:2012, Cholis:2012,
Rao:2012fh, Whiteson:2013cs, Carlson:2013vka}. Preliminary results from the
LAT team investigating the excess were presented in~\cite{bloom_charles_fermi_lat_line, Albert:Talk}.
 
Although the evidence for a 130 GeV line is compelling, significant concerns
about the Galactic center signature remain, including: (1) its statistical
robustness, and (2) its possible cause by an instrumental systematic. We will
briefly summarize the current situation, concentrating on statistical
properties of the Galactic center excess and -- as arguably the most worrisome
indication for a systematic -- a feature in the low incidence angle Earth
limb data. 
In the following, we will estimate the additional exposure needed to clarify
the situation on statistical grounds, and discuss a strategy to obtain it
while mitigating impact on other science objectives. 

\section{Current status of signal and systematics}
\subsection{Time evolution of the signal significance}
The first claims of a spectral feature \citep{Bringmann:2012,Weniger:2012}
were based on 3.5 years of LAT data (through 4 February 2012).  More recent data provide an opportunity to confirm the signal and see how it evolves with
time. Because the signal is based on approximately 1 signal count per month,
Poisson fluctuations cause the significance to accumulate in
something akin to a random walk.  The mean trend in significance builds with
the square root of exposure ($\mathcal{E}$) but with large uncertainties.  A confirmation at $3\sigma$ of the
signal in the original ROI (i.e.~no trials factor) would be persuasive, but
the exposure required to get an expected significance of $3\sigma$ is quite
different from the exposure required to achieve $3\sigma$ \emph{with a
probability of 95\%.}

\begin{figure}[h]
  \begin{center}
    \includegraphics[width=0.60\linewidth]{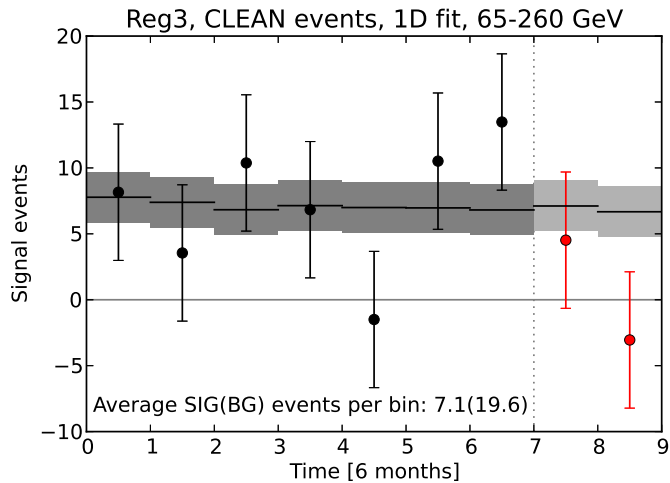}
    \vspace{-0.5cm}
  \end{center}
  \caption{Number of signal events in 6-month bins starting from 4
    August 2008, obtained by an unbinned likelihood fit to the data in
    Region~3 of Ref.~\cite{Weniger:2012}. The gray band shows the expected
    signal rate with $\pm1\sigma$ uncertainty as extracted from data taken until
    4 February 2012.
    In red we show data taken since 4 February 2012
    together with the projected event rate.}
  \label{fig:semester_fluxes}
\end{figure}

\begin{figure}[h]
  \begin{center}
    \includegraphics[width=0.6\linewidth]{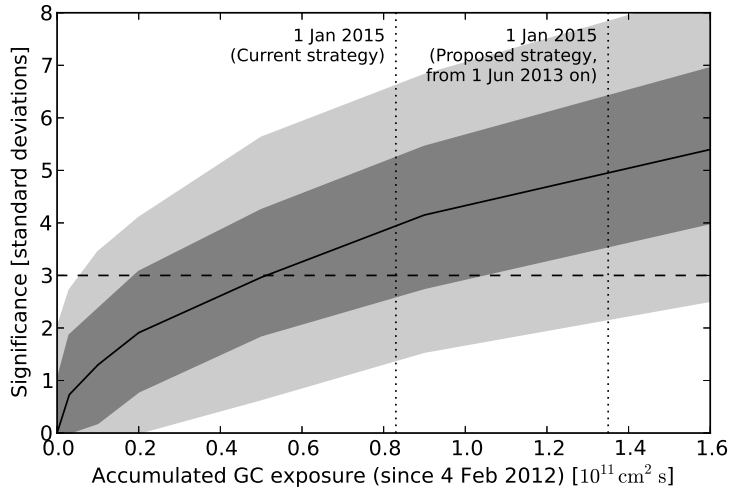}
    \vspace{-0.5cm}
  \end{center}
  \caption{
    Expected evolution of signal significance in Region~3 from
    Ref.~\cite{Weniger:2012}, starting on 4
    February 2012, as function of
    accumulated exposure.  The shaded bands show $68\%$ and $95\%$ CL
    uncertainties as derived from a Monte Carlo simulation.  The assumed
    signal rate is $1.2\pm 0.3$ events per month, the effective background is
    $3.3$ events per month (the values measured prior to 4 February 2012).
    The first and second vertical dotted lines indicate how much exposure is
    expected to be accumulated until end of 2014 if the observation strategy remains
    unchanged, and respectively when the observation strategy proposed in this
    document is adopted starting from June 2013. Note that without changing
    the survey strategy, until the end of the three year extension Aug 2016,
    the accumulated exposure will be $1.29\times10^{11}\rm\,cm^2\,s$.}
  \label{fig:projection}
\end{figure}

\begin{figure}[h]
  \begin{center}
    \includegraphics[width=0.6\linewidth]{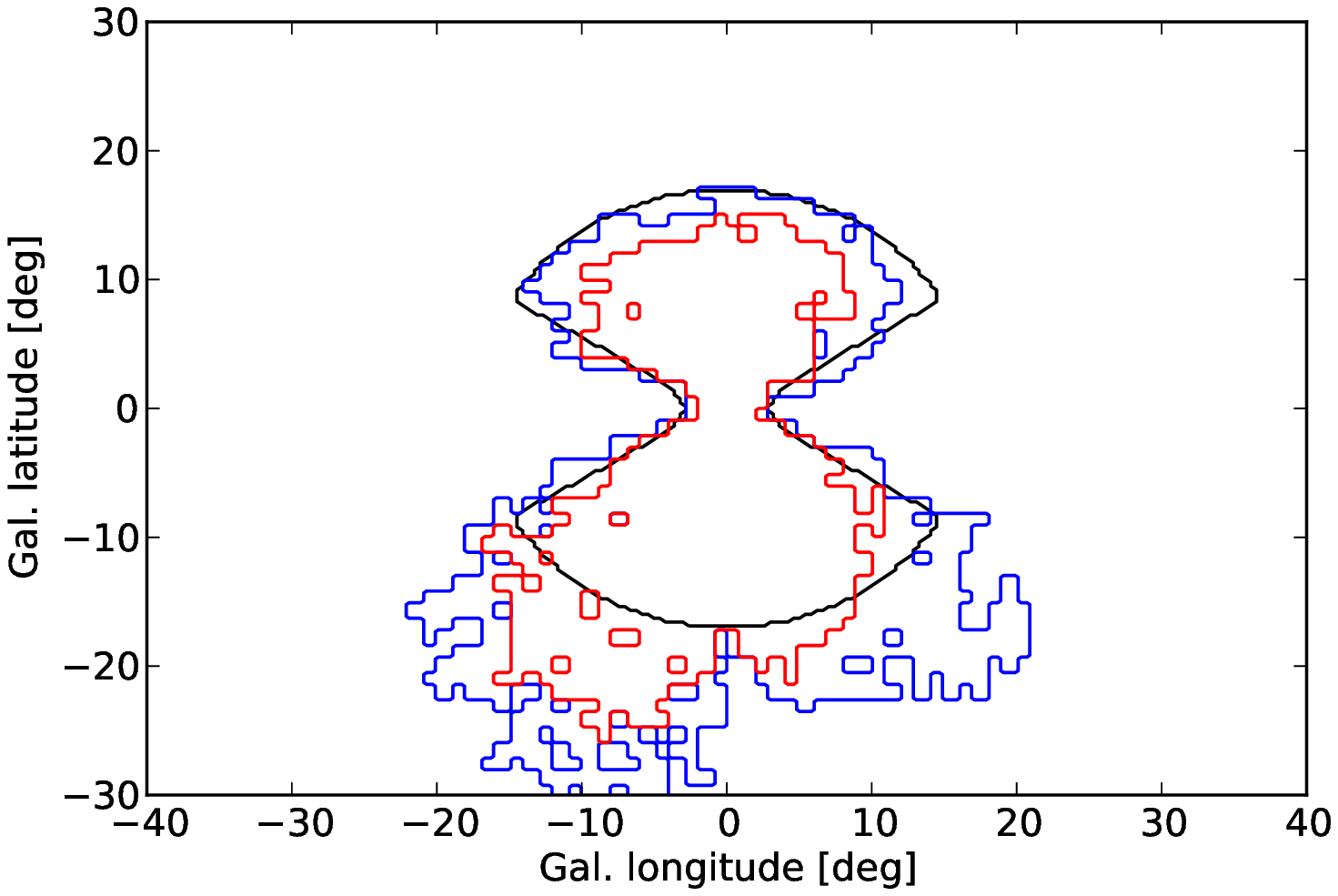}
    \vspace{-0.5cm}
  \end{center}
  \caption{Region 3 (blue) and Region 4 (red) used in
  Ref.~\cite{Weniger:2012}, together with the simpler hourglass region
  proposed for future studies (cp.~also with Fig.~3 in
  Ref.~\cite{Bringmann:2012ez}).}
  \label{fig:regions}
\end{figure}

\emph{Status.} 
We use the first 3.5 years of data, 4 August 2008 through 4 February 2012, to
define the spectral and spatial properties of our signal hypothesis (this is
the data set that was used in the initial publication,
Ref.~\cite{Bringmann:2012}). Data taken after the 4 February 2012 can then be
used to confirm or reject this hypothesis without trials. To be conservative,
we will not use the template regression technique from Ref.~\cite{linepaper},
but follow~\cite{Bringmann:2012, Weniger:2012} and use spectral fits in ROIs
with large expected S/N.  For definiteness, we will only use `Region 3' from
Ref.~\cite{Weniger:2012}, which is one of the regions where the excess was
first identified. Below we
propose a geometrically similar but much simpler region that we advocate as an
alternative for future studies.

We will here rely on the public unreprocessed Pass 7 data. Preliminary results
from the LAT team show that the line feature is also present in the
reprocessed Pass 7 data, though it moves to 135 GeV and appears to be slightly
less significant~\cite{Albert:Talk}. Throughout, we will refer to the feature
as `130~GeV feature' for simplicity.

We derive the number of excess events $N_s$ by a maximum likelihood fit to the
first 3.5 years of data from Region 3 (using a power-law plus line model). In
the fit, we adopt an energy range from 65 to 260 GeV, fix the line position at
$E_\gamma=129.8\rm\,GeV$, and use P7CLEAN events only.\footnote{The details of
the fit are identical what was done in Ref.~\cite{Weniger:2012}, but we
checked that similar results are obtained using Region 4, P7SOURCE events,
or a `2D' fit that takes into account incidence angle as well as energy
information in the modeling of the line.} We find a number of $N_s=50.0\pm 13.3$
excess events, with a
statistical significance of $s=4.3\sigma$. The effective number
of background events can be estimated as $N_b = N_s^2/s^2$, and is $N_b=137.4$.
These parameters together with the details of the fit define our \emph{signal
hypothesis}; the \emph{null hypothesis} is that the flux is compatible with a
single power law, i.e.~$N_s=0$.

In Fig.~\ref{fig:semester_fluxes} we show the distribution of these excess events in bins of six
months. The first seven bins correspond to the 3.5 years that define the
excess, the two red data points are data taken afterwards. Error bars include
background fluctuations and are given by $\Delta N_s = \sqrt{N_s+N_b}$. To
obtain $N_s$ and $N_b$, we fit the data of each six-month bin
individually. The gray bands show the number of signal events expected from
the above fit to the 3.5 years data, with small variations related to
the exposure.

The last two bins are compatible with both the null-hypothesis at the one
sigma level, and with the signal-hypothesis at the two sigma level (see also
Ref.~\cite{Weniger:2013dya}). In the case of a real signal, the last bin would
represent a decent downward fluctuation. However, it would be premature to
draw strong conclusions from this single observation.

\emph{Projection.} 
Going forward, the signal hypothesis predicts that the significance will build
as sqrt(exposure), but with large uncertainties.  It is essential to estimate
these uncertainties, because we need to know how much exposure is required to
\emph{cleanly} separate the signal from the null hypothesis. 

In Fig.~\ref{fig:projection} we show 68\% and 95\% CL bands for the projected evolution of the
signal significance as function of exposure accumulated after 4 February 2012.
To generate the plot, we simulate events from a power-law plus line model and
derive the signal significance for each realization as described above. The
background and signal correspond to the best-fit values obtained in Region 3
with data until 4 February 2012.  For different realizations we allow the
signal normalization to vary following a normal distribution that is matched
to the $\pm1\sigma$ flux uncertainties from the 3.5 year results.\footnote{We
checked that similar results for the projection evolution are obtained for
Region 4, in the hourglass region discussed below, or using the analytical
estimates from Ref.~\cite{Weniger:2013dya}.}

The first vertical dashed line indicates the exposure that will be acquired by
1 Jan 2015 with the current survey strategy (in the first 3.5 years, an
exposure of $10^{11}\rm\,cm^2\,s$ was collected at the GC at 100 GeV). It is likely that a true signal
would be confirmed with $>3\sigma$ significance, although this is not
guaranteed due to the $95\%$ CL error bands that range down to significances
of only $1\sigma$. Worse, it will be difficult to robustly rule out the signal
hypothesis at the $3\sigma$ level unless the accumulated significance is close
to zero.\footnote{To exclude the signal hypothesis at, say, the $2\sigma$ level at a
specific point in time, the actually observed significance has to lie outside
of the predicted
$95\%$ CL band in Fig.~\ref{fig:projection}.} This drives home the point
that, left unchanged, the current survey
strategy may well leave us with ambiguous results well into 2015.  As we will
discuss in the next section, a change in the observation strategy will allow a
much quicker confirmation/rejection of the signal hypothesis.
\medskip

\emph{Alternative ROI.} For future analysis of the 130 GeV feature in the LAT
data, instead of using the difficult to manage regions from
Ref.~\cite{Weniger:2012}, we propose to use the region shown in black in
Fig.~\ref{fig:regions}. This region is (1) geometrically simple and easy to reproduce, and (2)
it is centered on the part of the sky where the excess was largest in the
previous data. Though we avoid any statement about particular dark matter
profiles, it is an \emph{a priori} region for data taken since 4 February
2012. We defined this ROI in $(\ell, b)$ space as the intersection of
$\ell^2+b^2\leq
r^2$ and $\ell^2\leq (b\tan\varphi)^2 + d^2$, with $(r, d, \varphi) =
(20^\circ, 3^\circ, 60^\circ)$. 

\subsection{The Earth limb feature}
The continual cosmic-ray cascades in the Earth's atmosphere produce gamma rays
with $dN/dE \sim E^{-2.8}$~\citep{FermiLimb}. 
These so-called `Earth limb' photons provides a convenient source of photons
for systematics tests. 
The most direct indication for an instrumental cause of the Galactic center
feature is an excess of 130 GeV photons in low incidence angle events from
the Earth limb~\cite{linepaper, finkbeiner_systematics, Hektor:2012ev,
bloom_charles_fermi_lat_line}. Although it is challenging to understand how
such a feature could possibly be mapped onto the Galactic center while being
absent in other test regions, this feature has raised serious concerns about the
energy reconstruction of the LAT around 130 GeV.  Additional limb data would
determine whether the Earth limb feature is indeed a true systematic effect or
merely a statistical fluke in light of a large number of trials.  If it is a
reproducible systematic, additional data may be required to diagnose the
software or hardware problem responsible. 

We define here the Earth limb excess by selecting P7CLEAN events until 5 Sep
2012 at zenith angles $Z>100^\circ$,
incidence angles $\theta<60^\circ$, and excluding events within $20^\circ$ of
the Galactic center. In this data set, the line feature at 130~GeV has a
significance of $2.7\sigma$ when
fit in the range 65--260~GeV (for details of the
assumed line shape, etc.,
see Ref.~\cite{finkbeiner_systematics}).\footnote{If we take into account data
  until 16 Apr 2013, the
significance of the Earth limb feature is $3.0\sigma$, which is marginally
larger.}
Data accumulated after 5 Sep 2012 can
now be used to test whether this excess is spurious. Using the above
cuts, we find that the rate of events above 100 GeV between the
start of the mission and 5 Sep 2012 is 9.5 ph/month (in total 474), whereas
36.7 ph/month accumulated between 5 Sep 2012 and 16 Apr 2013. The reason for
this increase is the commencement of weekly dedicated limb observations as
well as two extended
target of opportunity observations in that time period. If the accumulation of
low incidence angle Earth limb data continues at this pace, 2.3 years of fresh
data will be enough to regenerate the putative limb signature with $4.0\sigma$
significance on average, or to rule it out with reasonably high significance if it is a
fluke. Without a change of the observation strategy, this amount of data would
be available end of 2014. As we will discuss next, a change of the
observation strategy would help to collect the same amount of low incidence
angle Earth limb data in a much shorter time period, and the additional data
would be invaluable for diagnosing instrumental problems.

\begin{figure}[t]
  \begin{center}
    \includegraphics[width=0.49\linewidth]{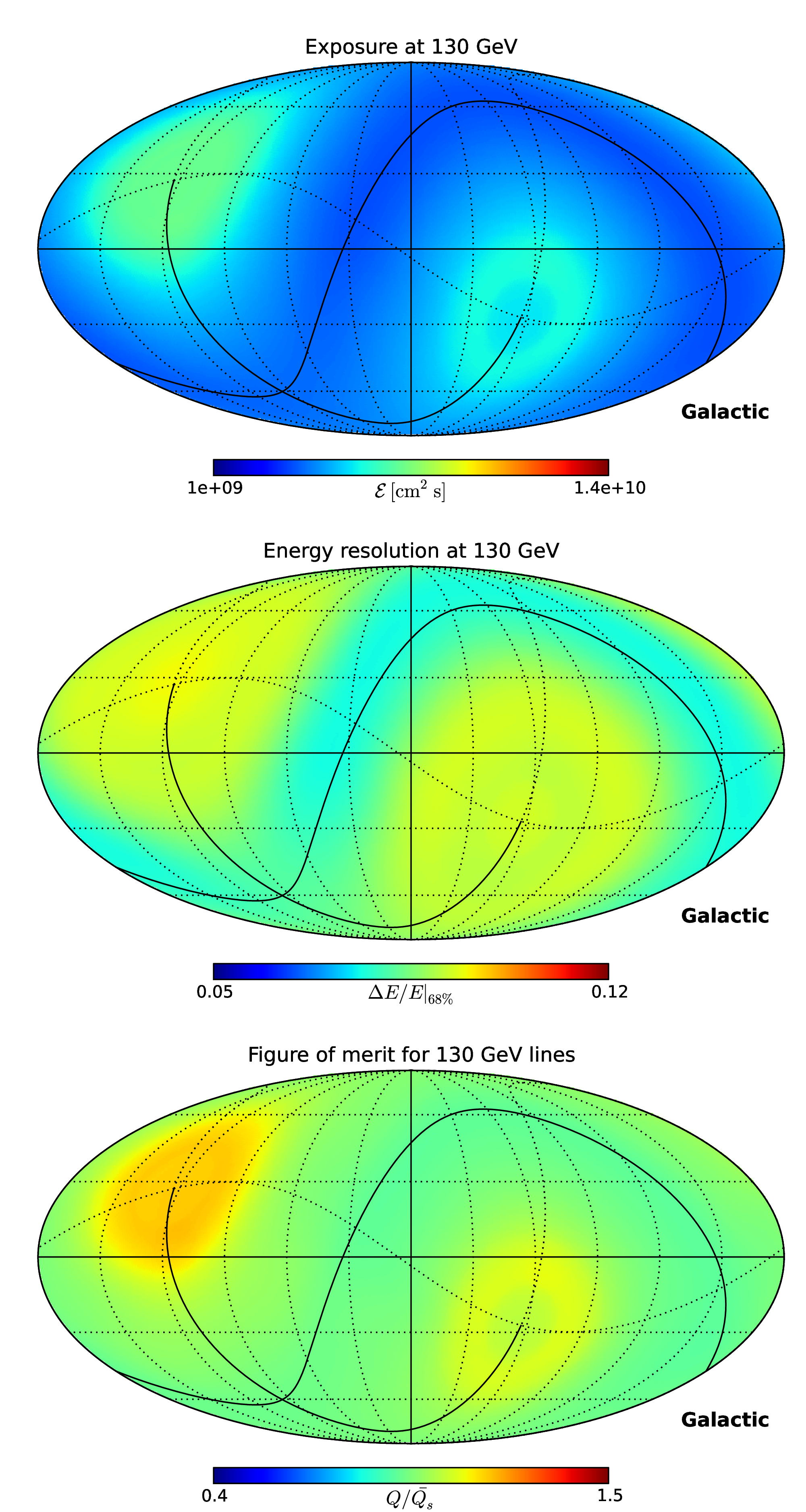}
    \includegraphics[width=0.49\linewidth]{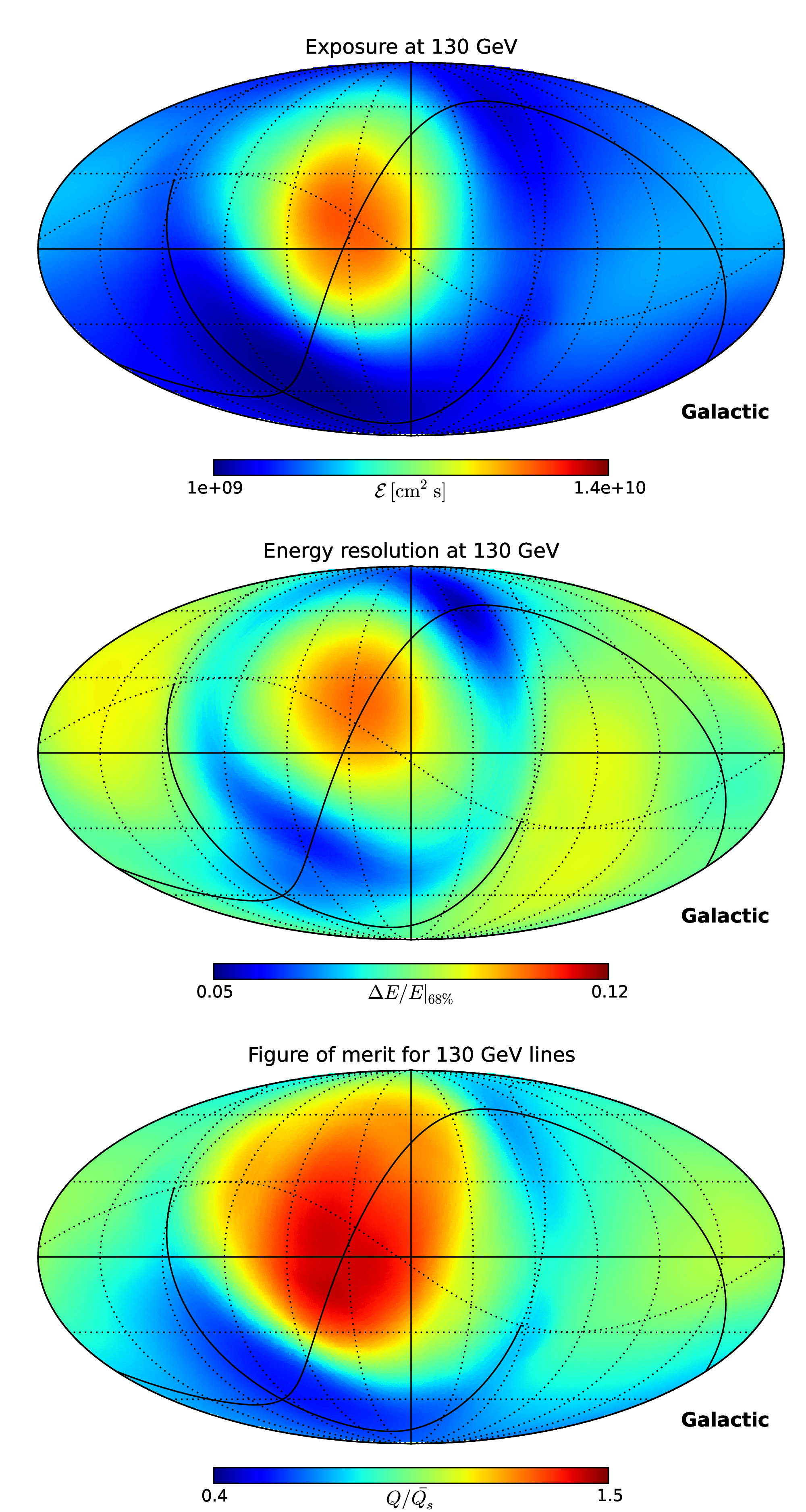}
    \vspace{-0.5cm}
  \end{center}
  \caption{Evaluation of standard survey mode (\emph{left panels}) and mixed observation
    strategy (\emph{right panels}; option3v3). Sky maps are in galactic coordinates ($\ell$ increases
    to the left) and averaged over a orbital precession period of 55 days.
    \emph{Top
      panels:} Exposure maps in cm$^2$s.
    \emph{Central panels:}
    Effective energy resolution (half 68\% containment width)
  \emph{Bottom panels:} Figure of merit for gamma-ray line searches. In all
  panels the overlaid lines show the main axes of the equatorial coordinate
  system; sky maps are symmetric around the celestial equator; color scales
  are linear.}
  \label{fig:mollweide}
\end{figure}

\begin{figure}[t]
  \begin{center}
    \includegraphics[width=0.49\linewidth]{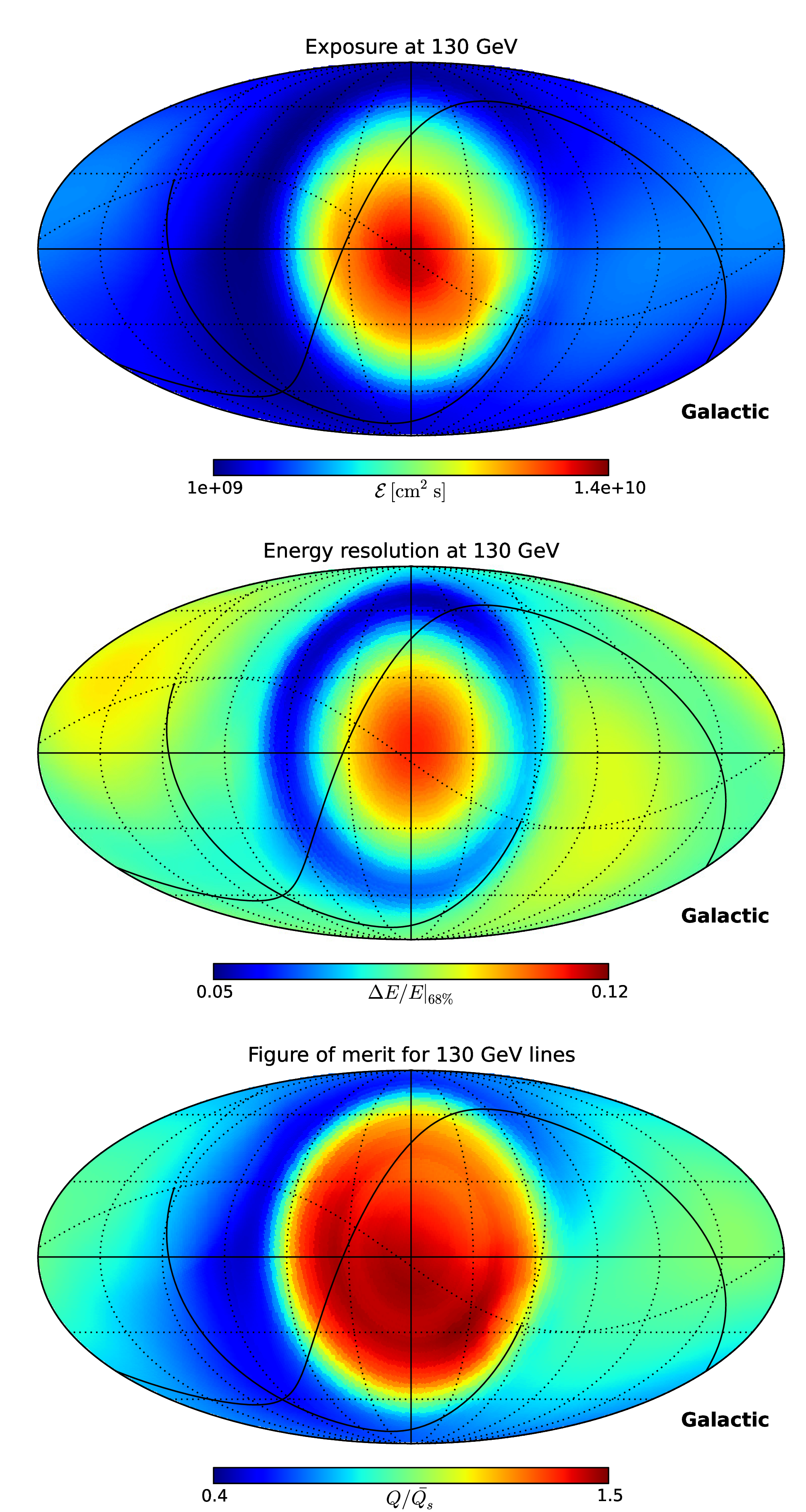}
    \includegraphics[width=0.49\linewidth]{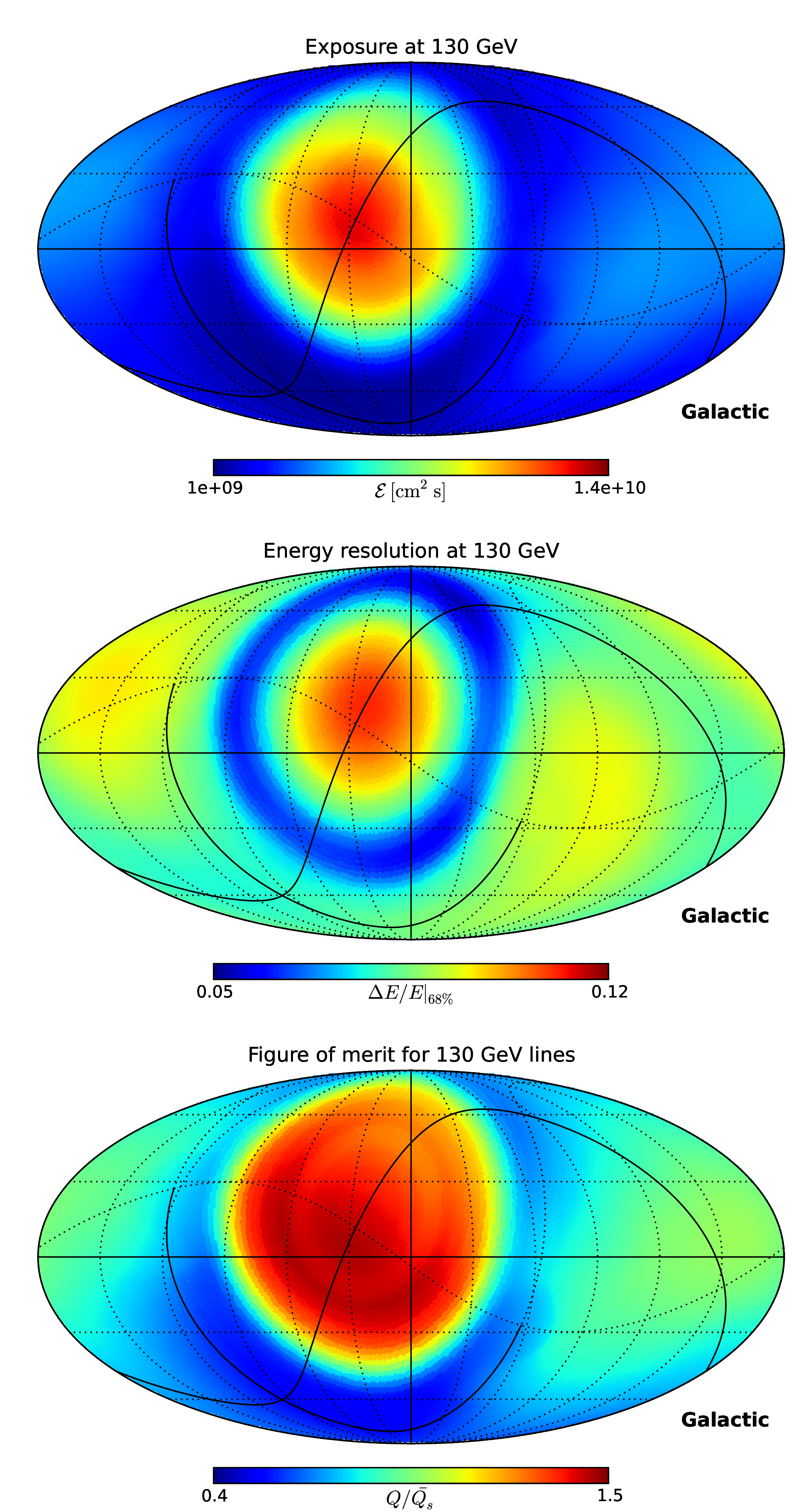}
    \vspace{-0.5cm}
  \end{center}
  \caption{Same as Fig.~\ref{fig:mollweide}, but for `option1v2' and
  `option2v2'.}
  \label{fig:mollweide2}
\end{figure}

\section{A New Observation Strategy}
Since the start of the mission, Fermi has spent over 95\% of the time in
\emph{standard survey mode}.
In this mode, the LAT points north of zenith towards the orbital pole by an
angle \zrock\ on one orbit, and south of zenith by the same angle on the next
orbit; the LAT pointing is confined to the plane perpendicular to its orbital
velocity. 
This observation profile, combined with the precession of the orbit every
$\sim53.4$ days, allows the LAT to observe the whole sky with approximately
uniform coverage. The standard survey mode is only occasionally interrupted
for pointed observations of targets of opportunity (ToOs). During such times
the LAT may point at a larger zenith angle than usual, even at the horizon.  In
addition, survey mode is occasionally interrupted by Autonomous Repoints of
the observatory for triggered gamma-ray burst follow-up observations, and for
calibration.

However, Fermi is capable of very flexible survey mode patterns. For example,
a single orbit may include both survey mode and pointed observations (``mixed
mode''), increasing coverage of certain parts of the sky. We will explore the
impact of such a strategy on the study of the 130 GeV feature at the Galactic
center.  We focus here on the mixed modes `option1v2', `option2v2' and
`option3v3' put forth for discussion by the Fermi mission.\footnote{See
\url{http://fermi.gsfc.nasa.gov/ssc/proposals/alt_obs/obs_modes.html}.}

Our goal is to increase the exposure on the Galactic center.  The
basic strategy would be to switch to pointed observation of the Galactic
center when possible, and to follow the standard survey mode otherwise. More
precisely, \Fermi\ would slew from survey mode to the target once the target
is $10^\circ$ from Earth occultation, and slew back to survey mode position
once the target reenters $10^\circ$ from Earth occultation.
The Earth Avoidance Angle (EAA) is set to
$30^\circ$, to avoid the loss of too much exposure during the transition
periods. This means that the LAT will track the target only to within $30^\circ$ of
the Earth limb and then hold steady before it switches back to survey mode. 

To reduce potential systematics, it is advisable to avoid pointing directly at
the target. Instead, it is useful to observe it with a broad distribution of
incidence angles. The three mixed modes option1v2, option2v2
and option3v3 differ mainly in what target position exactly is adopted.
In option1v2 and option2v2, the position is fixed at (RA,
Dec)=($261.4^\circ$, $-28.9^\circ$) and (RA, Dec)=($261.4^\circ$, $0^\circ$),
respectively.\footnote{The Galactic center is at RA=266.4$^\circ$,
Dec=-28.9$^\circ$.}  In option3v3 the target position RA is set to
261.4$^\circ$, while the target declination varies within the
range Dec=$\pm25.6^\circ$ during one orbital precession period such that the
target position remains close to the orbital equator.  These variations of the target
position yield an improved sky uniformity on short time scales. 

\subsection{Impact on line searches at the Galactic center}
The upper panels of Fig.~\ref{fig:mollweide} show exposure maps after 55 days
survey mode (left) and mixed mode option3v3 (right) in galactic
coordinates.  In the mixed mode, the point of highest exposure is at (RA,
Dec)$\simeq(261.4^\circ$, $0^\circ)$. At the Galactic center, the exposure increases by a
factor of 2.23 relative to normal survey mode.

\begin{figure}[t]
  \begin{center}
    \includegraphics[width=0.30\linewidth, angle=-90]{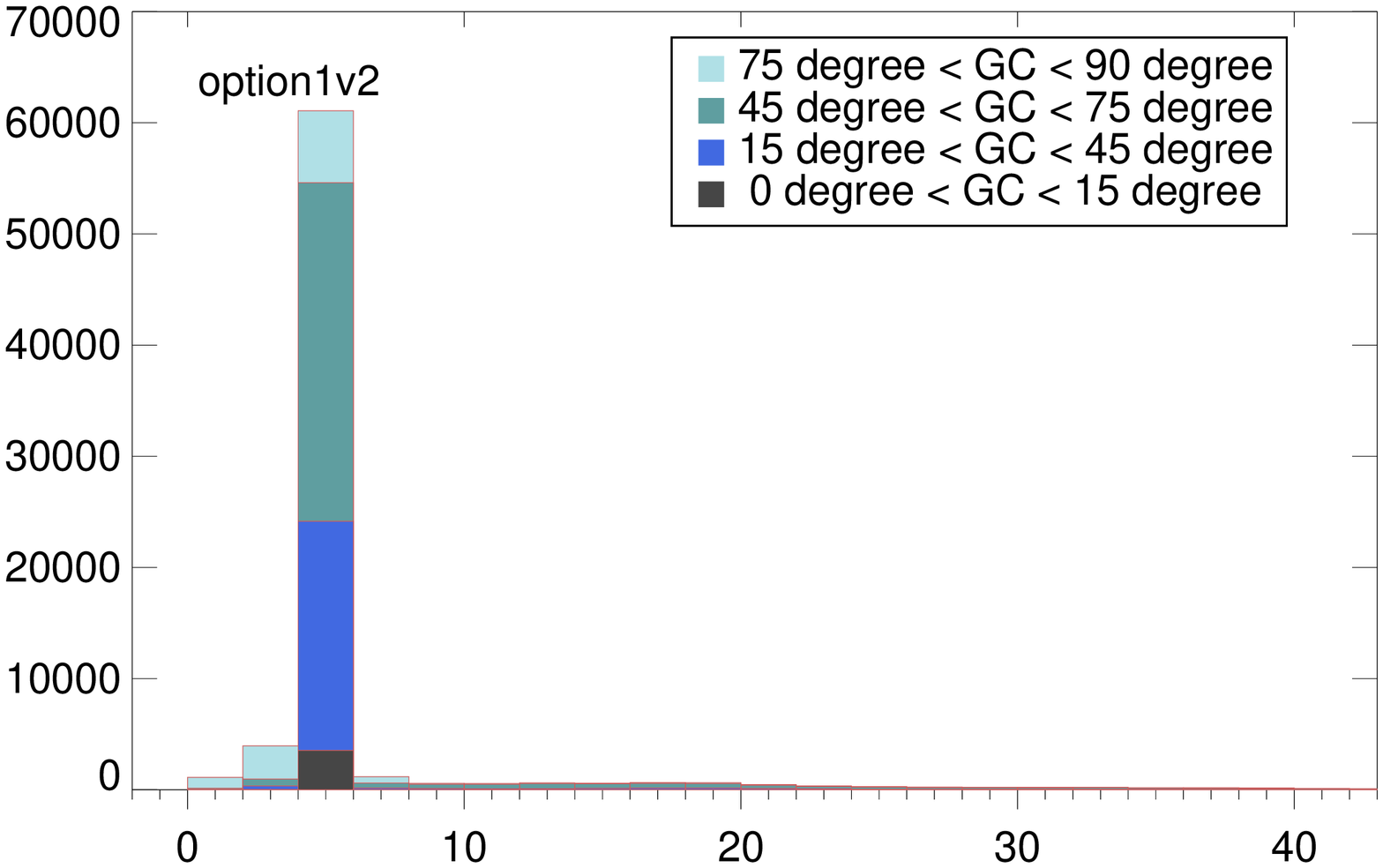}
    \includegraphics[width=0.30\linewidth, angle=-90]{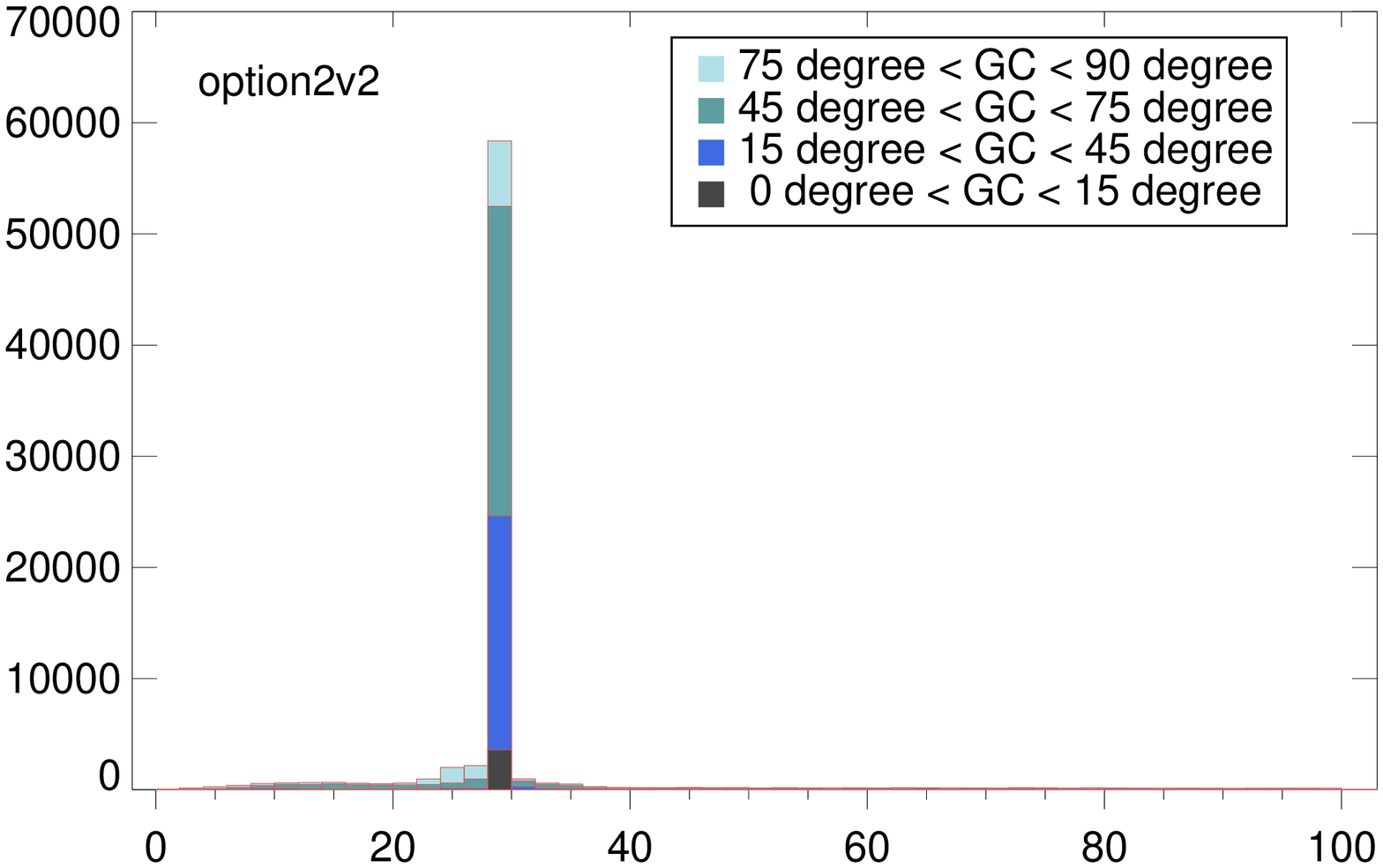}
    \includegraphics[width=0.30\linewidth, angle=-90]{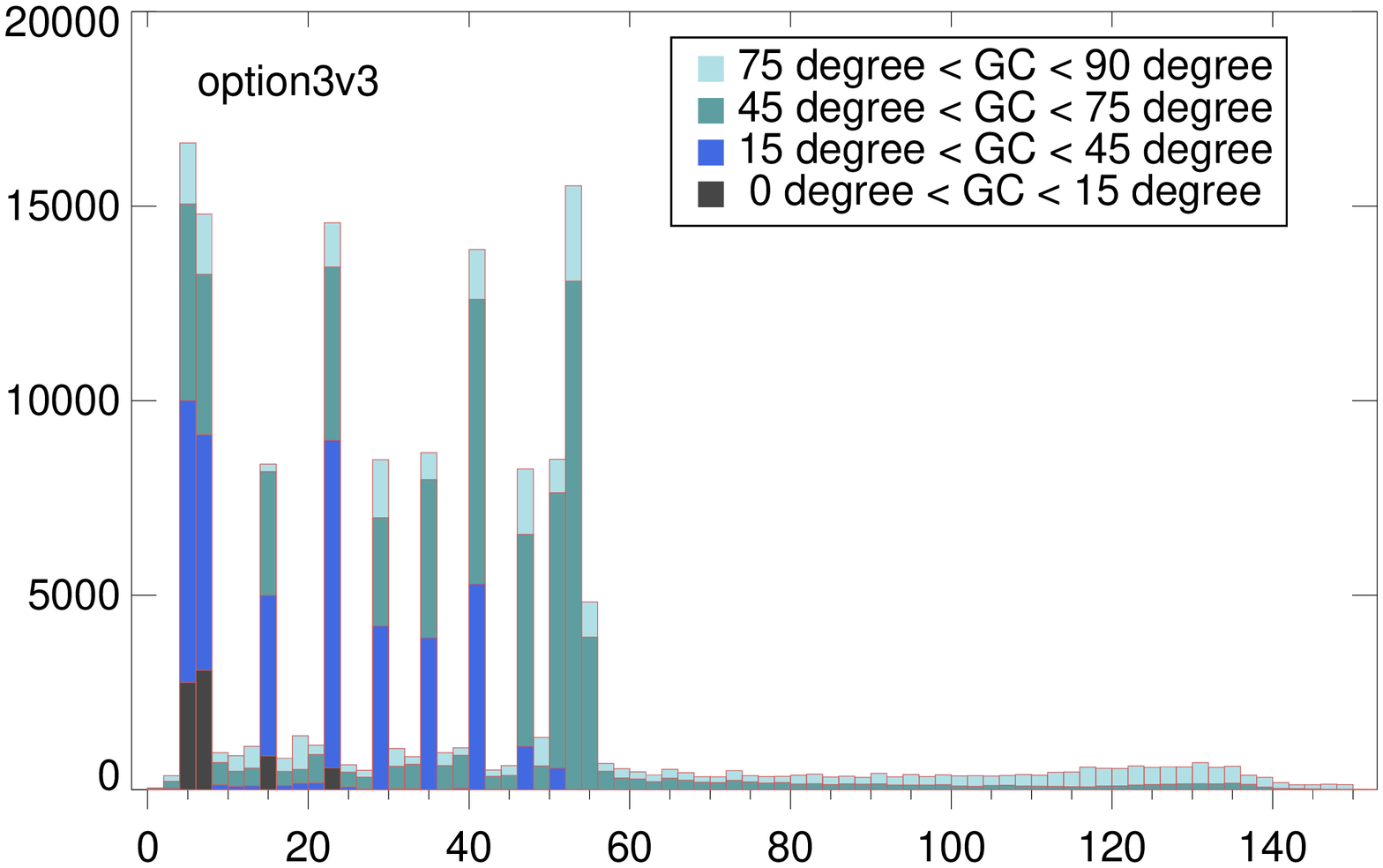}
    \includegraphics[width=0.30\linewidth, angle=-90]{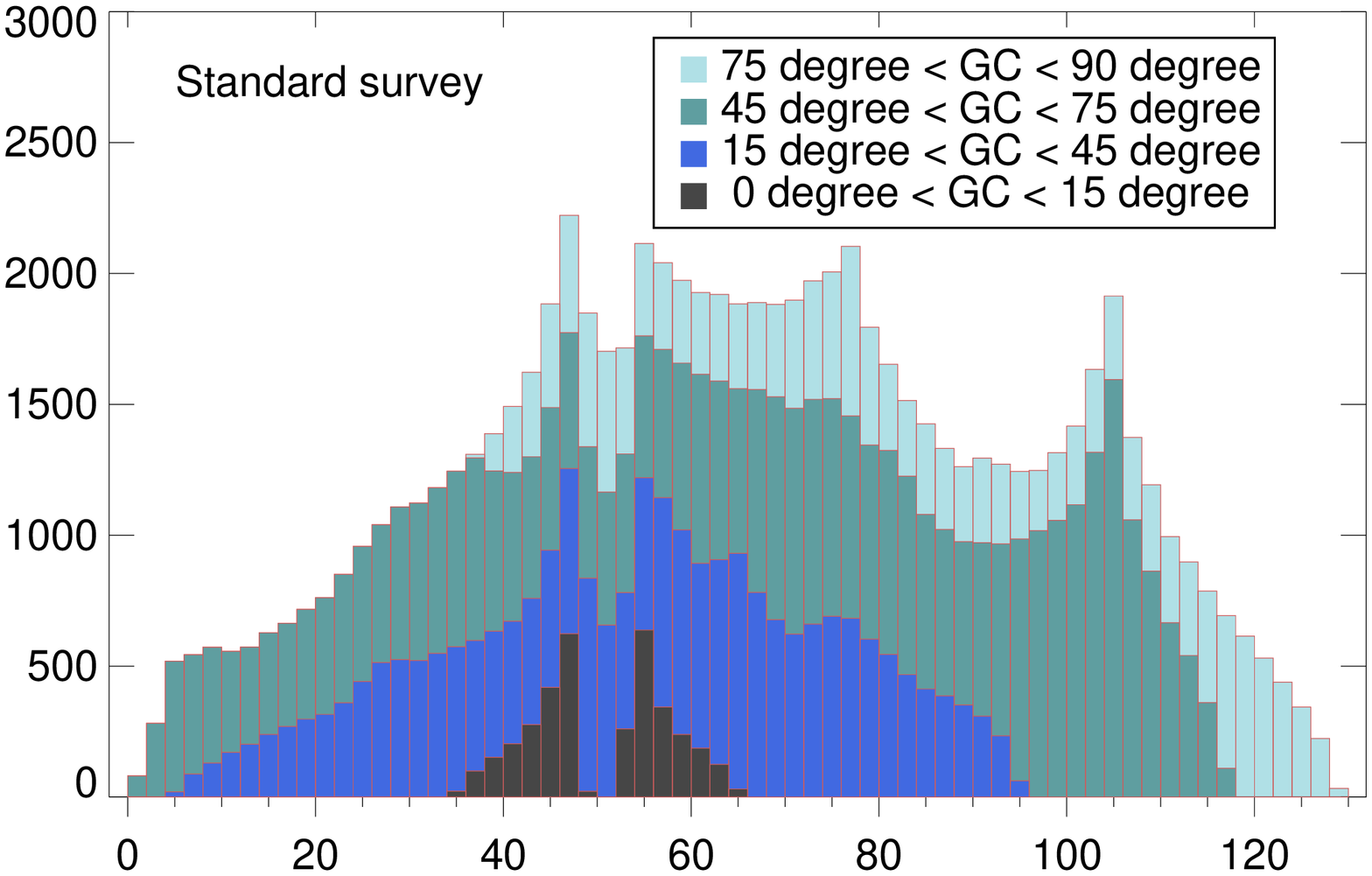}
    \vspace{-0.5cm}
  \end{center}
  \caption{The distribution of Galactic center theta angle. Each entry in the
  histogram corresponds to a 30 seconds interval. The colors correspond to
  the zenith angle of the Galactic center. The plots take account of time spent in the SAA.} 
  \label{fig:thetaDist}
\end{figure}

In mixed mode, regions close to the Galactic center are predominantly observed
at low incidence angles in the range $\theta\simeq5^\circ$--$50^\circ$ (see
Fig.~\ref{fig:thetaDist}). This has impact on the effective energy resolution,
which is shown in the central panels of Fig.~\ref{fig:mollweide}. In direction
of the Galactic center the energy resolution is in fact slightly worsen with respect to the
standard survey mode ($\Delta E/E=9.59\%$ for option3v3 instead of $\Delta E/E=8.75\%$).
However, this loss of resolution has only a small effect on line searches.

In Fig.~\ref{fig:mollweide2}, we show the same information, but for the mixed
modes option1v2 (left) and option2v2 (right). The Galactic center exposure increases by a
factor 3.0 and 2.5, respectively, whereas the effective energy resolution is
$11.1\%$ and $10.0\%$. Apart from the characteristics at the Galactic center, these modes
mainly differ in the impact on other science goals, as we will discuss below.
\medskip

As a convenient \emph{figure of merit} for line searches we define the
dimensionless quantity $$Q\equiv a\sqrt{\mathcal{E}/\Delta E}\,,$$ which is
proportional to the expected median line significance in units of standard
deviations.  Here, $\mathcal{E}$ is the exposure in cm$^2$s, $\Delta E$ the
energy resolution, and $a$
normalizes $Q$ such that the spatial mean in survey mode is $\bar Q=1$. In the
bottom panels of Fig.~\ref{fig:mollweide} and~\ref{fig:mollweide2} we show sky maps for $Q$ in mixed and
survey mode.  At the Galactic center, $Q$ increases by a factor $1.43$ when
switching to mixed mode option3v3, which would increase the growth rate of
the signal significance by $43\%$, equivalent to doubling the exposure/time. 

\subsection{Impact on Earth limb observations}
In general, to keep systematics that might be related to specific incidence
angles $\theta$ under control, it is useful if the Galactic center is observed
at a broad distribution of different incidence angles. For the four
observation modes, this distribution is shown in
Fig.~\ref{fig:thetaDist}. In case of option3v3, $\theta$ spans a range from
$5^\circ$ to $50^\circ$ (the discrete distribution comes from jumps in the
target position as it follows $\rm Dec\simeq0^\circ$). This is a significant
advantage
when compared to option1v2 and option2v2, which feature pronounced peaks at
$\theta\sim5^\circ$ and $\theta\sim30^\circ$, respectively.

An important side effect of the mixed mode is the accumulation of
additional Earth limb data at low incidence angles, which can be used for
checks of instrumental systematics. This happens during the
transitions between pointed observation and survey mode. The target comes
close to the horizon, while the satellite maintains a minimal distance of
$30^\circ$ from the Earth limb. Consequently, the Earth limb is observed at
incidence angles $\theta\gtrsim30^\circ$ twice every orbit (1.5 hours).

\begin{figure}[t]
  \begin{center}
    \includegraphics[width=0.6\linewidth]{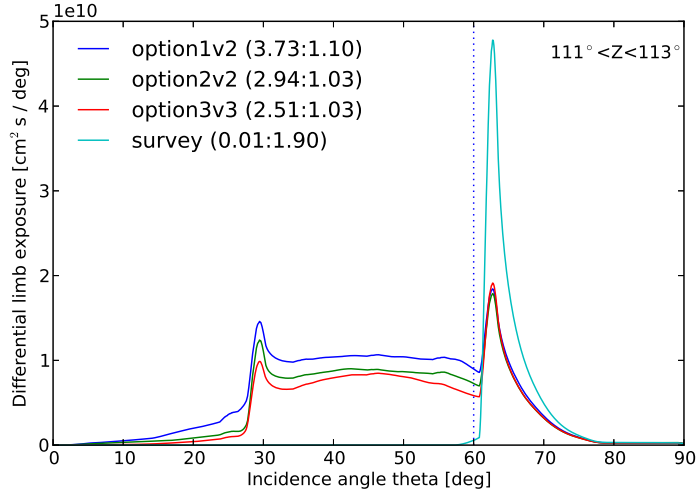}
    \vspace{-0.5cm}
  \end{center}
  \caption{Differential exposure of Earth limb region (zenith angles
    $111^\circ<Z<113^\circ$) as function of incidence angle $\theta$. In
    standard survey mode, incidence angles below $\theta<60^\circ$ are not
    exposed to the Earth limb. However, in mixed observation modes, a
    large number of Earth limb events would be collected down to incidence
    angles $\theta\simeq30^\circ$. In brackets we show the exposure below and
  above $\theta=60^\circ$ in units of $10^{11}\rm\,cm^2\,s$.}
  \label{fig:limb_exposure}
\end{figure}

In Fig.~\ref{fig:limb_exposure}, we show the expected differential exposure of
the Earth limb at zenith angles $111^\circ<Z<113^\circ$. In standard survey
mode, practically no limb data is collected at $\theta\lesssim60^\circ$.
However, during mixed mode a very significant number of the detected Earth
limb events would be collected at lower incidence angles. This would
accelerate the accumulation of low incidence angle Earth limb data with
respect to the previous 4.5 years by about \emph{a factor of five}, and will
allow a rapid rejection or confirmation of the 130 GeV feature in part of the
Earth limb spectrum. Note that a collection of Earth limb events at incidence
angles even lower than $30^\circ$ is possible if the EAA is set appropriately,
however at the cost of losing more exposure on the rest of the sky.

\subsection{Impact on other science}
Changing from survey mode to mixed mode observation will necessarily drag
exposure from some parts of the sky towards the Galactic center. In order not
to
lower the scientific power of Fermi, it is imperative that the
\emph{integrated} exposure over the whole sky remain as high as
possible. Furthermore, for the observation of transient phenomena, it is vital
that all parts of the sky are sufficiently covered at least each day; `blind
spots' should be avoided.

\begin{table}[t]
  \begin{tabular}{lcrcr}
    \hline
    Mode && Mean exposure [$10^9\rm\,cm^2\,s]$ && GC exposure
    [$10^9\rm\,cm^2\,s]$ \\\hline
    survey && 4.74 && 4.33 \\
    option1v2 && 4.38 && 13.1 \\
    option2v2 && 4.44 && 10.7 \\
    option3v3 && 4.56 && 9.67  \\
    \hline
  \end{tabular}
  \caption{Mean exposure (averaged over full
  sky) and exposure of Galactic center for different observation profiles in comparison.
  We assume P7CLEAN events at 100 GeV and 55 days of observation.}
  \label{tab:exposures}
\end{table}

In Table~\ref{tab:exposures}, we compare for the four reference 
strategies the mean sky exposure obtained after 55 days of observation. The
overall loss in sky exposure w.r.t.~to the standard survey mode is small and
between $4\%$ (option3v3) to $8\%$ (option1v2).

\begin{figure}[t]
  \begin{center}
    \includegraphics[width=0.39\linewidth, angle=-90]{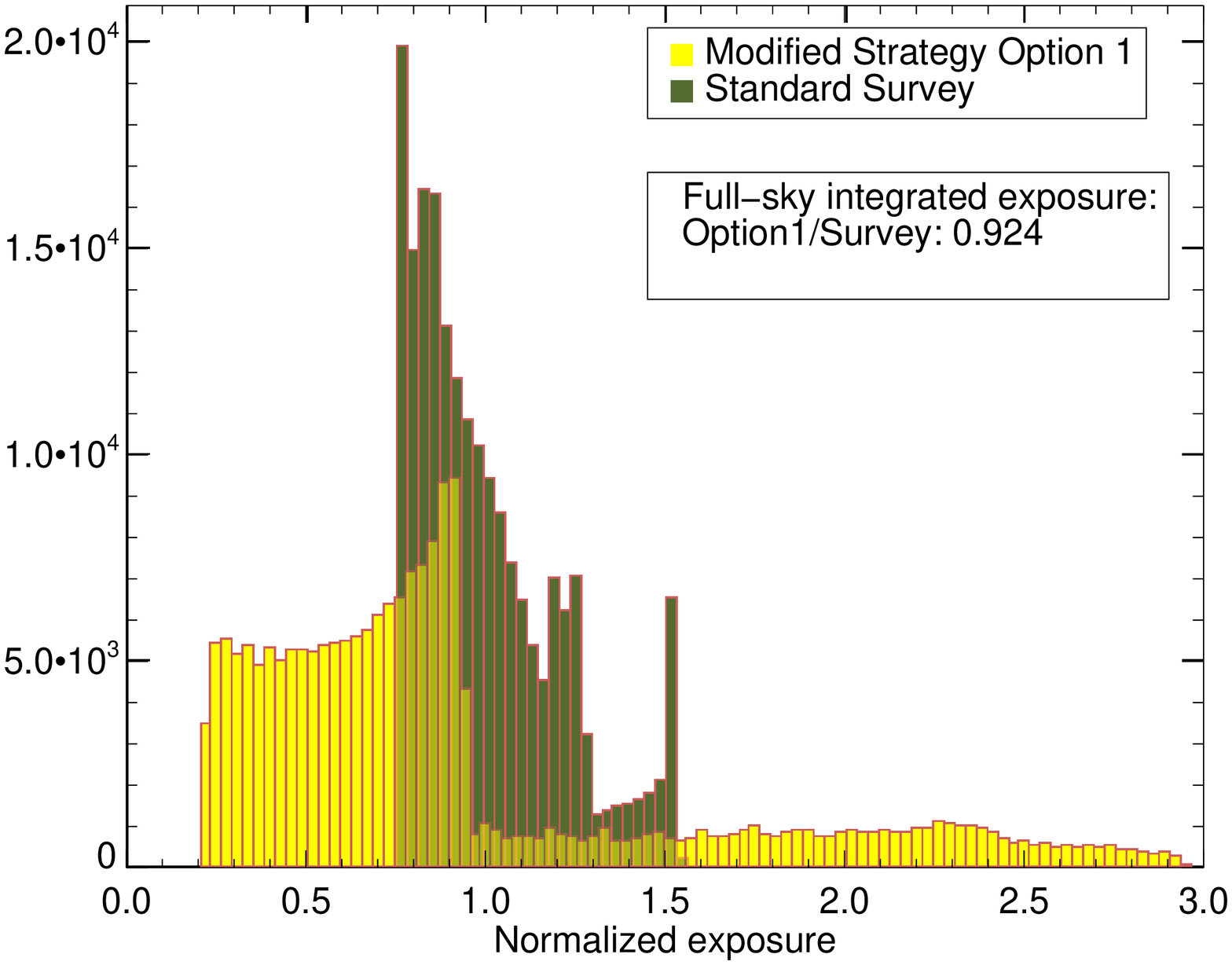}
    \includegraphics[width=0.39\linewidth, angle=-90]{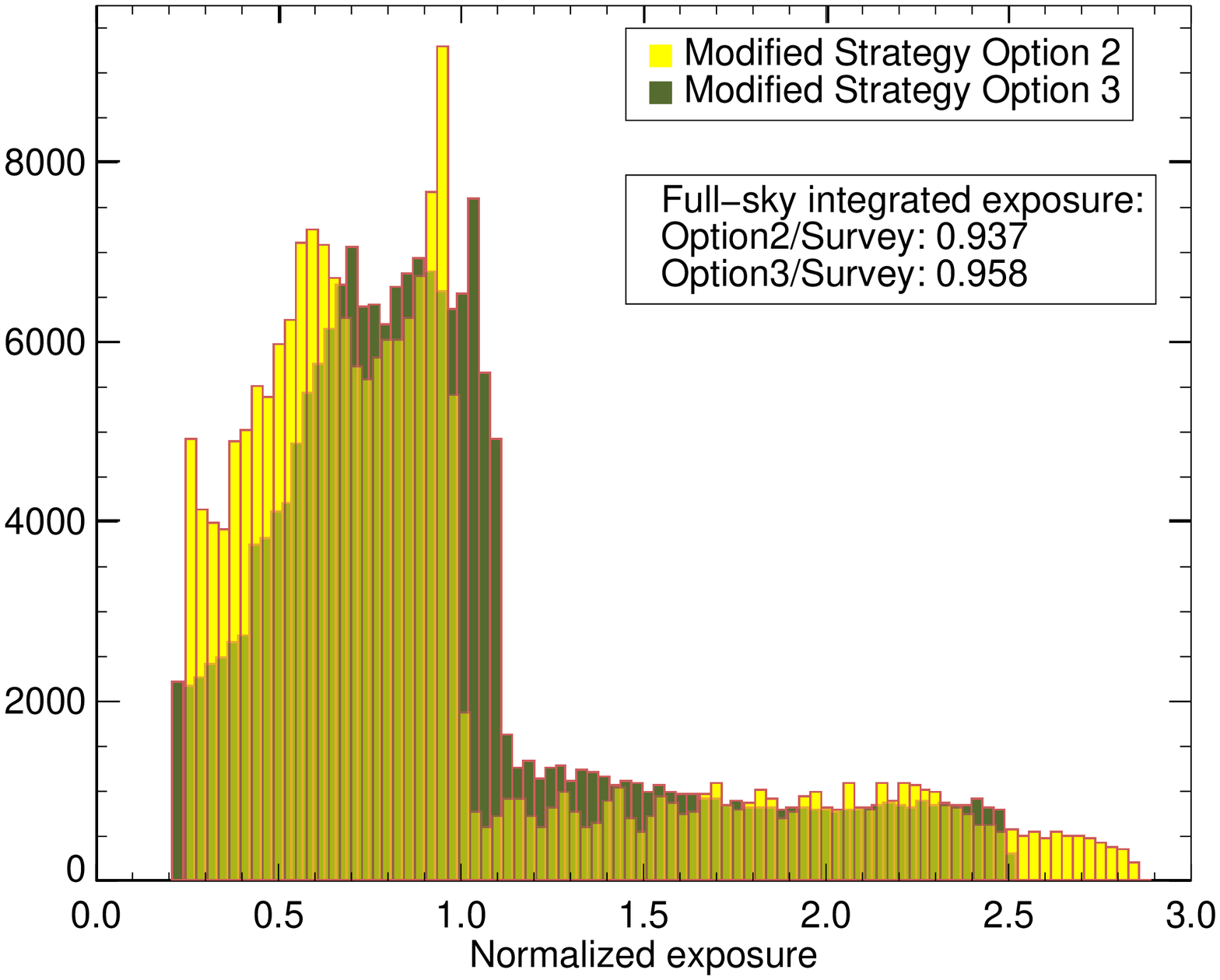}
    \vspace{-0.5cm}
  \end{center}
  \caption{Histogram of exposure per sky pixel, normalized to standard survey
    strategy. Left panel compares the
    mixed mode option1v2 to the standard survey strategy, and the right panel
    shows the modified strategy option2v2 and option3v3.}
  \label{fig:expHisto}
\end{figure}

The variations of the exposure over different regions of the sky are
illustrated in in Fig.~\ref{fig:expHisto}, where we show a histogram of the
distribution of exposure in different sky pixels (using a Healpix projection
with $N=128$). In the case of standard survey mode, the exposure distribution
spans a factor of two (being largest at Dec=$\pm90^\circ$), whereas it
spans a factor $10$ in the mixed mode option3v3. However, in no region of the
sky does the exposure drop by more than a factor of four relative to normal
survey mode.

\begin{figure}[t]
  \begin{center}
    \includegraphics[width=0.38\linewidth, angle=-90]{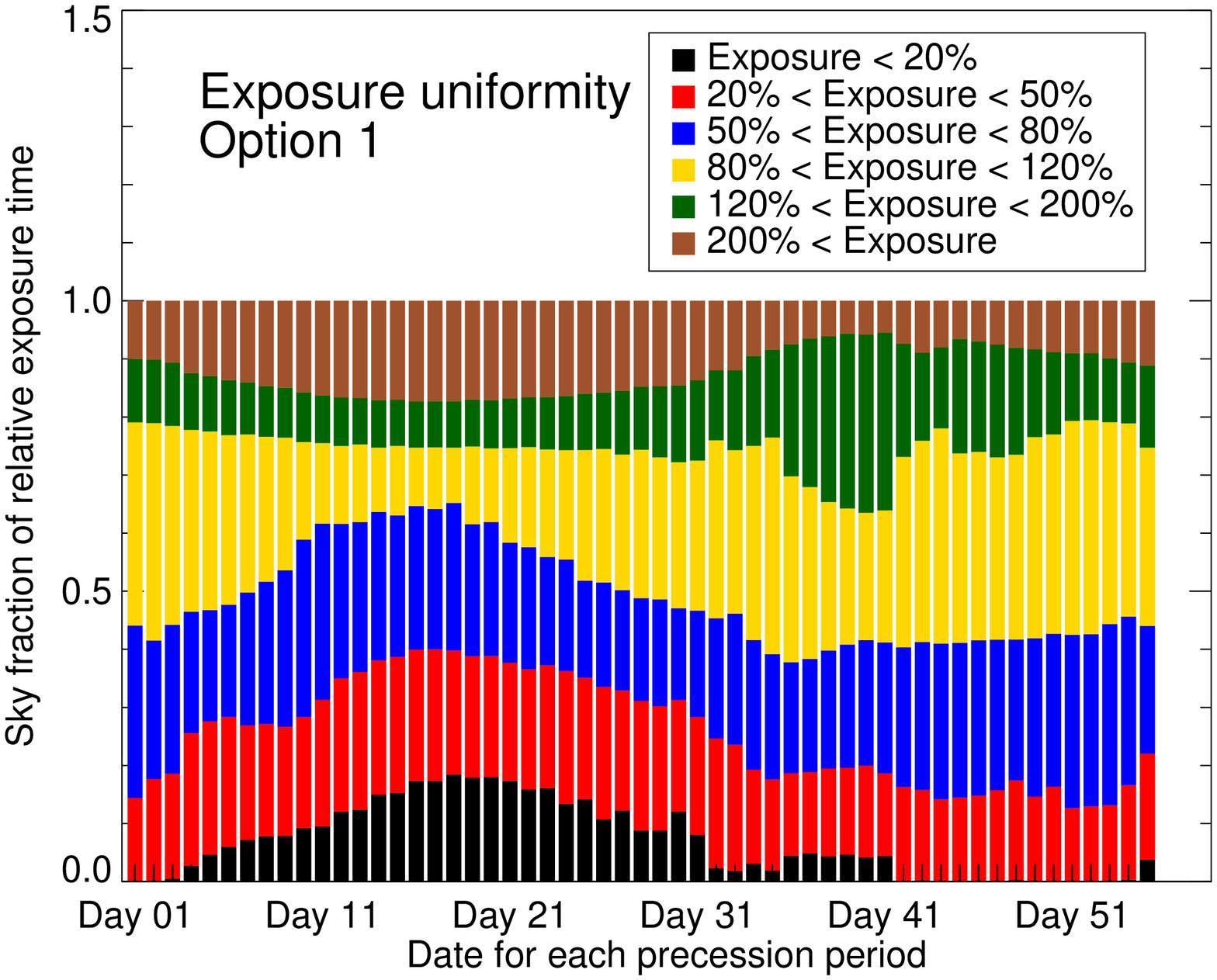}
    \includegraphics[width=0.38\linewidth, angle=-90]{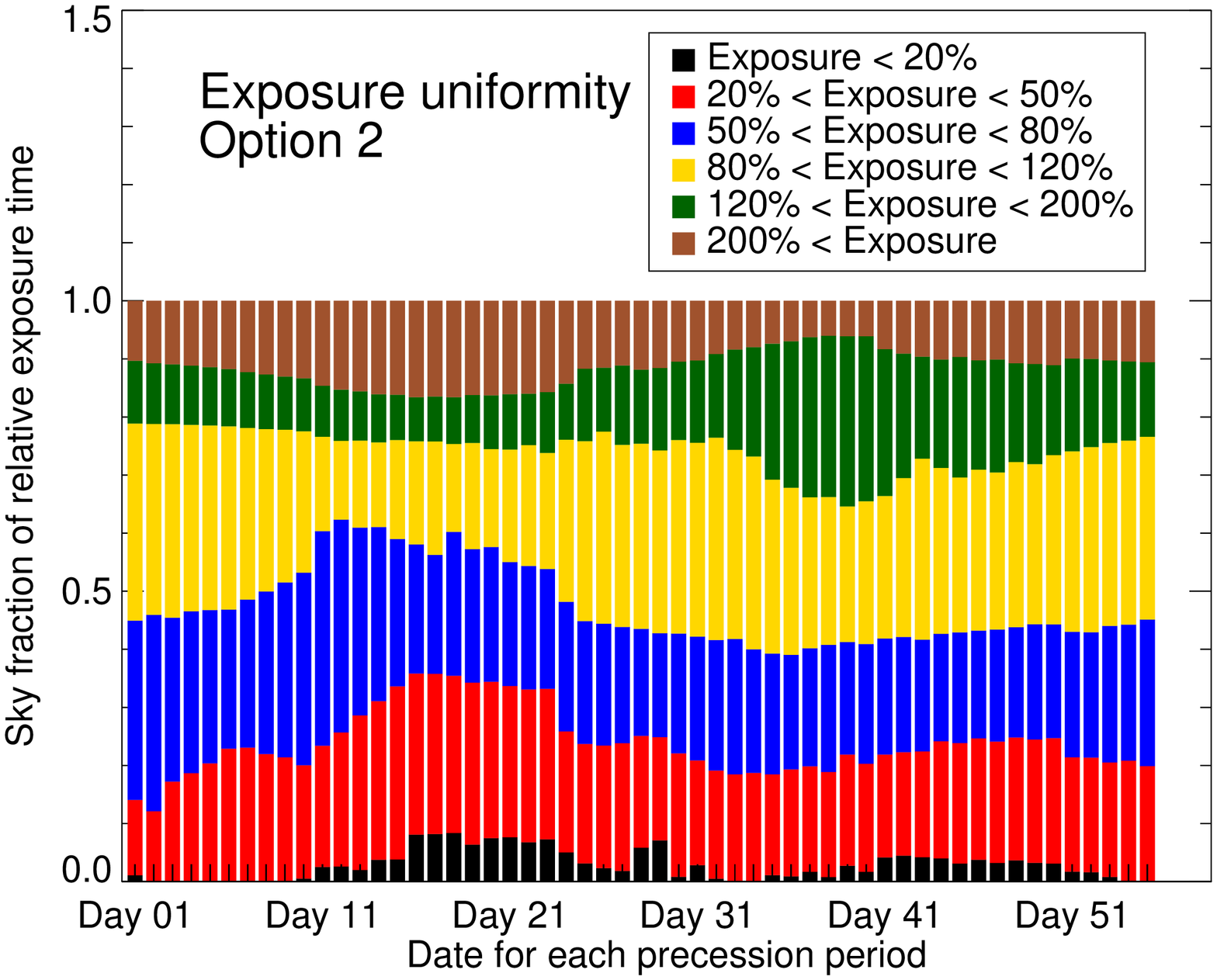}
    \includegraphics[width=0.38\linewidth, angle=-90]{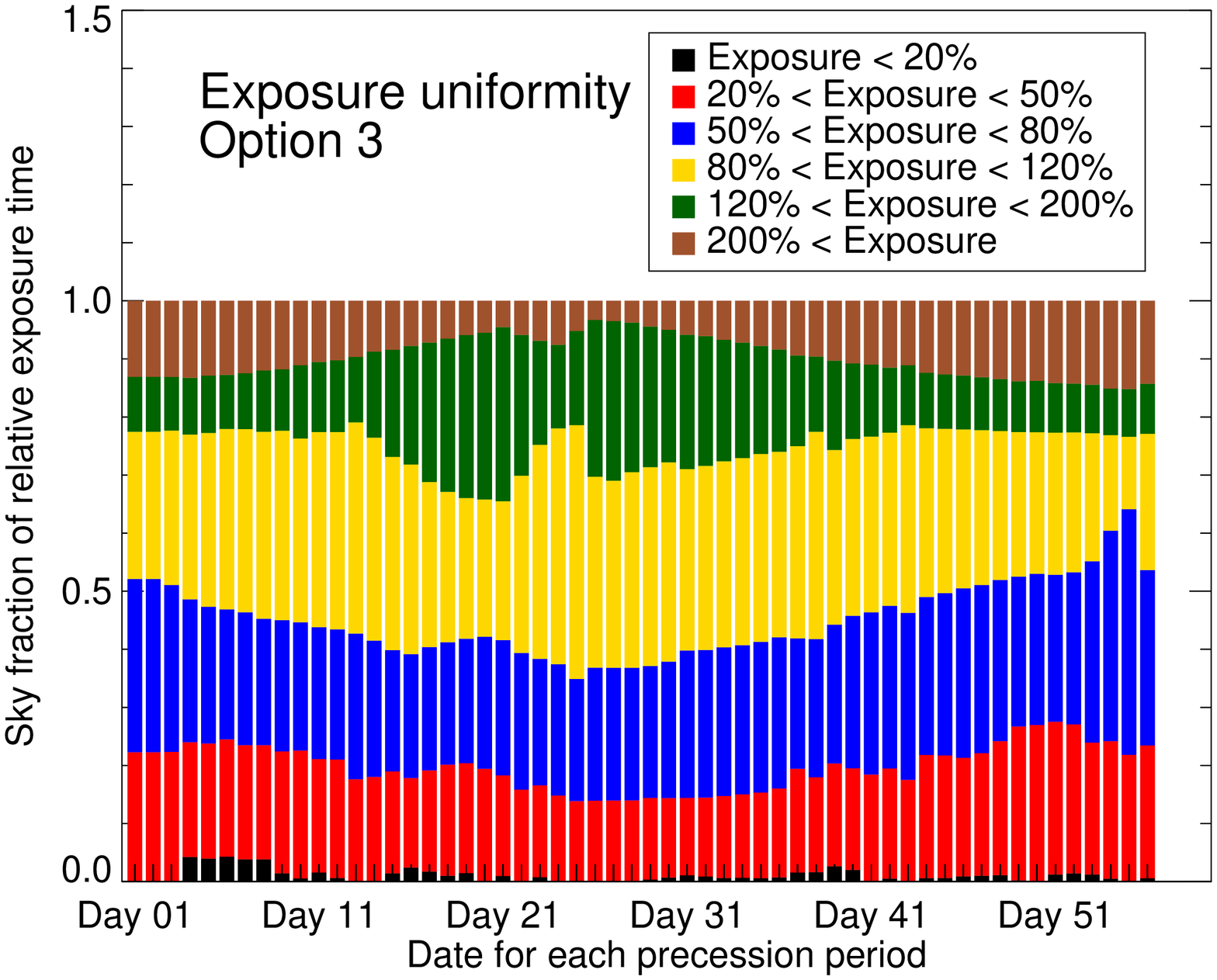}
    \includegraphics[width=0.38\linewidth, angle=-90]{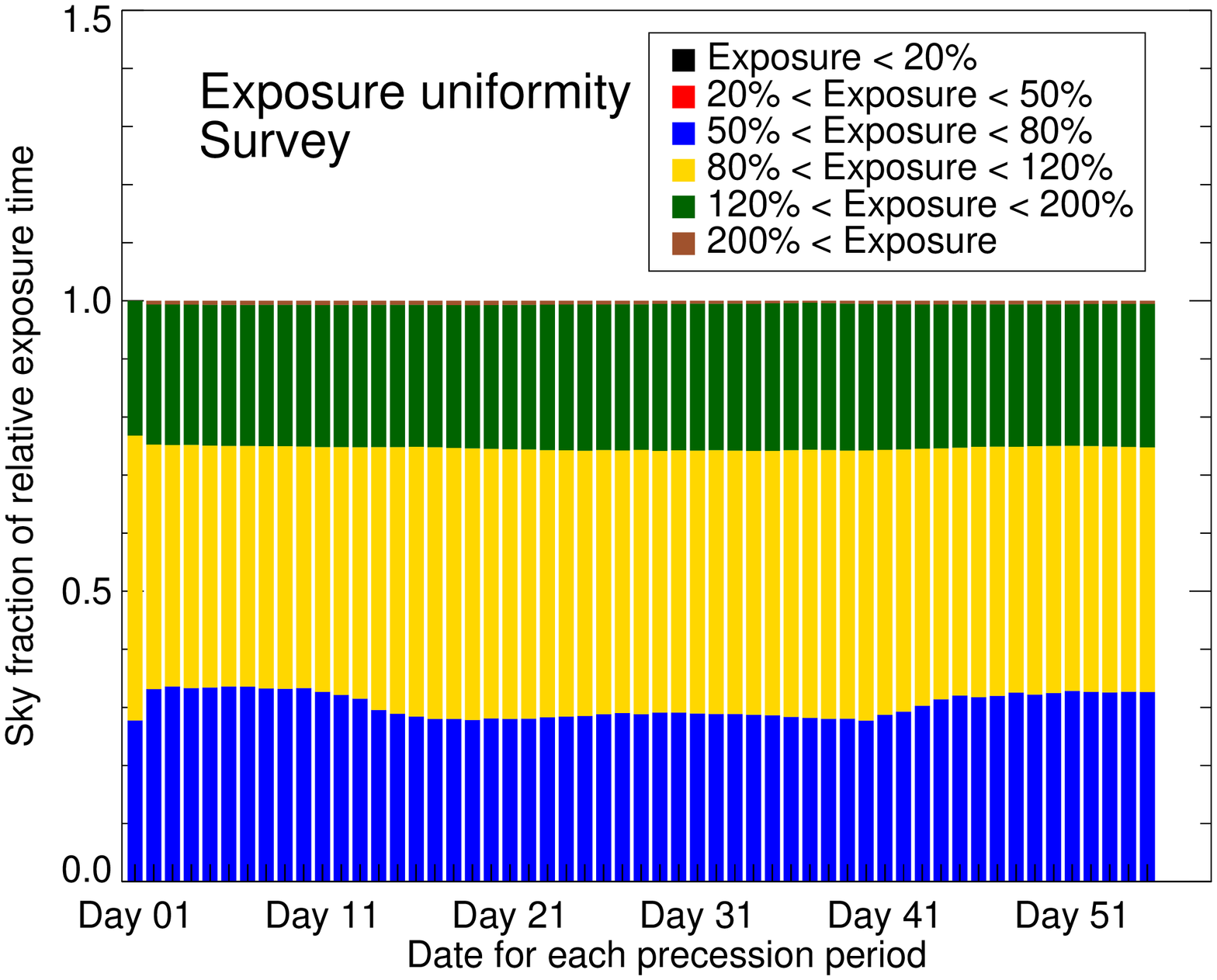}
    \vspace{-0.5cm}
  \end{center}
  \caption{Daily sky coverage with different range of exposure
  time normalized to the mean value of the exposure map of each day (in total
  55 days to complete on orbit precession period).} 
  \label{fig:coverage}
\end{figure}

For transient phenomena, the daily sky exposure is of importance. In 
Fig.~\ref{fig:coverage} we show for each individual day of the precession
period which fraction of the sky is covered by what fraction of the mean
exposure. In case of the mixed mode observations option2v2 and option3v3, in
less than $5\%$ of the sky the daily coverage drops below $20\%$ of the daily
mean, whereas in $>80\%$ the coverage remains above $50\%$ of the mean.

\begin{figure}[t]
  \begin{center}
    \includegraphics[width=0.5\linewidth, angle=-90]{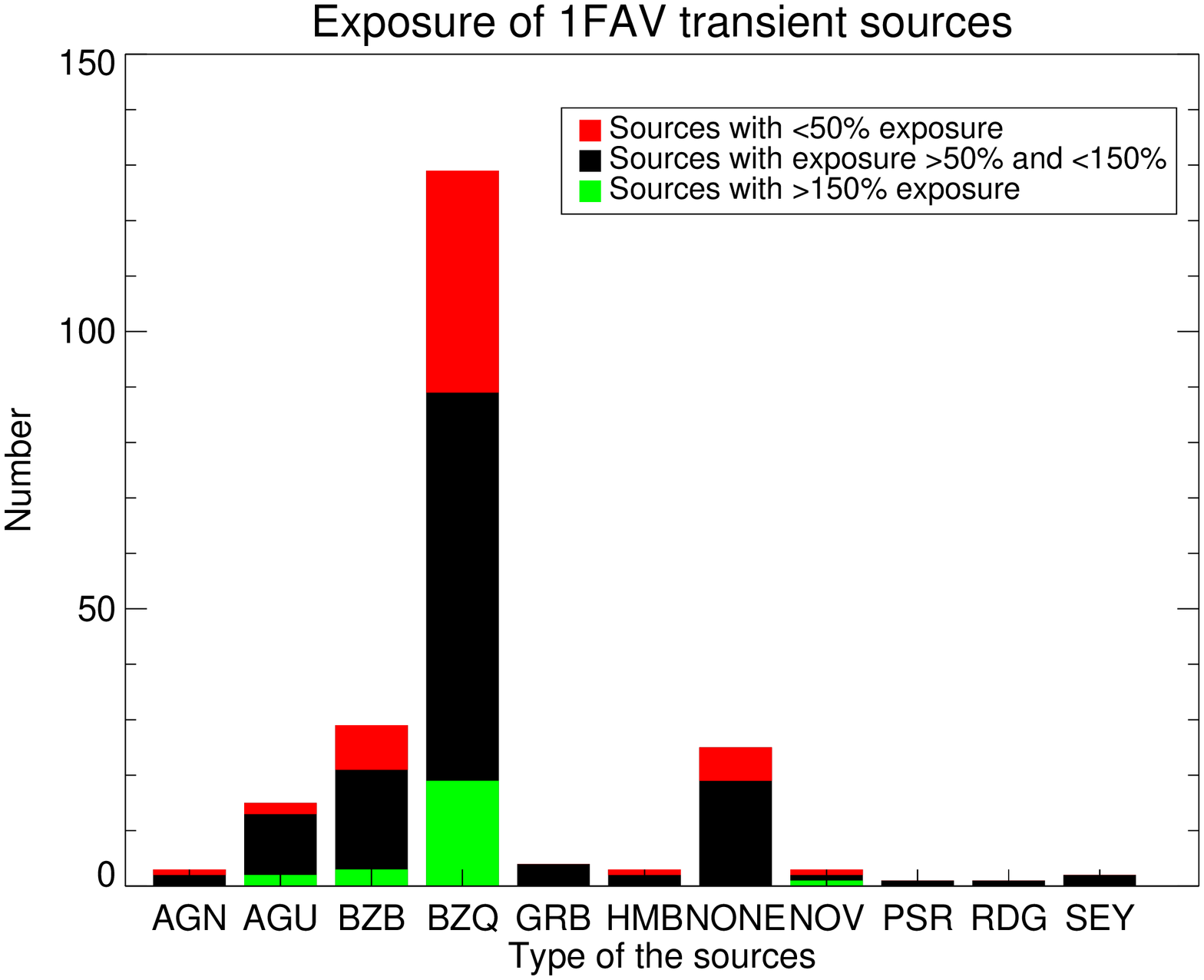}
    \vspace{-0.5cm}
  \end{center}
  \caption{Source type distribution for the 1FAV catalog
    from Fermi All-sky Variability
    Analysis~\cite{Fermi-LAT:2013jsa}. We classify the
    sources into three categories: sources with expected
    exposure time (assuming option1v2) less than 50\% of the
    standard survey strategy; sources with expected exposure
    between 50\% to 150\% compared to the standard survey
    strategy; and sources with expected exposure more than
    150\%. }
  \label{fig:transhist}
\end{figure}

\begin{table}[t]
  \begin{tabular}{lcrcrcr}
    \hline
    Mode && $>150\%$ exposure && $<50\%$ exposure && $50\%<$ exposure $<150\%$\\
    \hline
    option1v2 && 25 ($12\%$) && 59 ($27\%$) && 131 ($61\%$) \\
    option2v2 && 32 ($15\%$) && 45 ($21\%$) && 138 ($64\%$) \\
    option3v3 && 29 ($13\%$) && 30 ($14\%$) && 156 ($73\%$)\\
    \hline
  \end{tabular}
  \caption{Number (fraction) of transient sources corresponding to different expected exposure given three modified survey strategy. The fraction of expected exposure is compared to the standard survey strategy. }
  \label{tab:transsource}
\end{table}

The Fermi All-sky Variability Analysis has detected a total
of 215 flaring gamma-ray sources over the entire sky based
on weekly time intervals of 47 months LAT
data~\cite{Fermi-LAT:2013jsa}. In Fig.~\ref{fig:transhist}
we show how the exposure of different variable source types
could be affected by changing from survey to mixed mode
observations. For the different source classes, we show the
fraction of sources that would lose in exposure by more than
$50\%$ (red), which would gain more than $50\%$ in exposure
(green), and which remain relatively unaffected (black). For
option1v2, we have 25 sources that would receive at least
150\% exposure compared to the standard survey strategy with
finer sampling on time domain (including the recently
detected Nova Sco 2012), although with the price that 59
sources would receive less than 50\% exposure. We note that
61\% of the transient sources (131 in total) would receive
comparable (50-150\%) exposure time for option1v2 and the
standard survey mode.  In Table~\ref{tab:transsource}, we
compare for the three modified strategies the fraction of
transient sources that would obtain $>150\%$ exposure,
$<50\%$ exposure, and between 50-150$\%$ exposure time. As
we see from the Table, option3v3 provides a comparable
number of sources with $>150\%$ exposure, but a 50\% (33\%)
smaller sample of sources with $<50\%$ exposure for
option1v2 (option2v2).

For transient searches on different time scales, both
short-term flares on time scales of a few hours and
long-term flux variations of a few months, the changed survey
strategy would provide the opportunity for a sample of sources
with better sampling in the time domain and higher sensitivity
of flare detection.

Furthermore, the recently discovered G2 cloud - the intriguing red
emission-line object which is quickly approaching Sgr A$^*$ - offers a unique
opportunity to observe an accretion event onto the Galaxy's central black
hole~\cite{Gillessen:2011aa, Phifer:2013taa}. Although the predicted
intensity of gamma-ray emission from the G2 cloud encounter is not clear yet,
Fermi as the only gamma-ray space telescope covering from MeV to TeV energy
range, should produce the most detailed monitoring before, during and after
the closest approach of G2 to the central black hole with enhanced sensitivity
as a bonus of the modified survey strategy. Fermi's observation would provide
unique/key information to study this very rare event. The predicted encounter
time is about later this year, a change of survey strategy as soon as
technically possible would be suggested as response to this valuable chance
nature provides to us.

\subsection{Possible triggers}
We advocate the change to a new survey strategy as soon as technically
feasible. A more conservative approach might involve waiting for certain
triggers before initiating a change. For example, one might want to wait until
the Pass 8 processing is finished before making a decision.  However, it will
probably not be public for another year or more, and it is difficult to tie
the decision to an internal release; in order for the community to provide
input, the decision should be made based upon public data. 

Another trigger might be to just wait until a certain significance against/in
favour of the signal hypothesis is achieved. However, this could cause a
long-enough delay to make a later change inefficient. It would make more sense
to make the change immediately, and then have a trigger to revert to normal
survey mode when the signal hypothesis is ruled out at some level.

\section{Discussion and Conclusion}
\label{sec:Conclusion}

In this document we have argued for a change in the Fermi survey strategy to
increase exposure in the inner Galaxy, and confirm or rule out claims of a 130
GeV spectral line.  The principal reasons for a change are:

\begin{itemize}

\item{\it It is important: discovery of a dark matter annihilation line in the
    Galactic center would be Fermi's greatest accomplishment.}  The nature of
  dark matter is one of the greatest mysteries in physics and astrophysics,
  and the discovery of a line would be a major step forward for both fields.
  Exploring the nature of dark matter is one of the major goals of the Fermi
  project, and a discovery would define Fermi's legacy.

\item{\it Fermi can do it: a modified survey strategy can obtain a decisive
  measurement, while the status quo may not.}  The significance of a line
  evolves as sqrt(exposure), but with large uncertainty due to Poisson
  fluctuations.  For example, if the signal hypothesis is correct, and
  starting a new `trial-free' measurement from February 2012, the
  expected signal significance by 1 Jan 2015 is about $4\sigma$ (Fig.
  \ref{fig:projection}). However, Poisson fluctuations and uncertainties in
  the signal strength broaden this range such that the actual significance
  achieved is between 2.5 and 5$\sigma$ 68\% of the time, and between 1.5 and
  6.5$\sigma$ 95\% of the time. Hence a clear confirmation with $>3\sigma$
  significance is by no means guaranteed. The same is true for the rejection
  of the signal hypothesis if it is false. If the project continues with standard
  survey mode until 2015, there is a fair chance that we leave this question
  unresolved. We cannot permit this to happen.

\item{\it This is a win-win: the proposed change is not bad for other science.}  There will be winners and
  losers in any change, but more time on the inner Galaxy is good for lots of
  projects (better time coverage for pulsars and transients, etc.).  Many
  wide-angle surveys (SDSS, Pan-STARRs, etc.) have found it fruitful to
  dedicate a significant fraction of observing time to ``deep fields'' where
  greater sensitivity and improved cadence extend the range of phenomena
  observable.  Furthermore, roughly half
  the sky has more exposure under the new strategy, and even the underexposed
  regions are still observed on a regular basis for continued monitoring of
  transients (Fig. \ref{fig:coverage}).

\end{itemize}

For all of these reasons, we advocate a change in the survey strategy, as soon
as technically feasible. For reasons discussed above, `option3v3' put forward by the Fermi
mission seems to be the best way to go.

At least for the next several years, Fermi is uniquely able to address the 130
GeV line (with possible competition from HESS-II~\cite{Bergstrom:2012vd}).
If it is an artifact, it is a subtle one -- and understanding its
origin is important for the dark matter search in
particular, and the mission as a whole. If the line is real, we would forever
regret missing this opportunity to pursue it aggressively. 

\section*{Acknowledgments}
MS and CW thank the \emph{Kavli Institute for Theoretical Physics} in Santa
Barbara, California, for kind hospitality during the final stages of this
work.

\clearpage
\bibliography{whitepaper}

\begin{thebibliography}{47}%
\makeatletter
\providecommand \@ifxundefined [1]{%
 \@ifx{#1\undefined}
}%
\providecommand \@ifnum [1]{%
 \ifnum #1\expandafter \@firstoftwo
 \else \expandafter \@secondoftwo
 \fi
}%
\providecommand \@ifx [1]{%
 \ifx #1\expandafter \@firstoftwo
 \else \expandafter \@secondoftwo
 \fi
}%
\providecommand \natexlab [1]{#1}%
\providecommand \enquote  [1]{``#1''}%
\providecommand \bibnamefont  [1]{#1}%
\providecommand \bibfnamefont [1]{#1}%
\providecommand \citenamefont [1]{#1}%
\providecommand \href@noop [0]{\@secondoftwo}%
\providecommand \href [0]{\begingroup \@sanitize@url \@href}%
\providecommand \@href[1]{\@@startlink{#1}\@@href}%
\providecommand \@@href[1]{\endgroup#1\@@endlink}%
\providecommand \@sanitize@url [0]{\catcode `\\12\catcode `\$12\catcode
  `\&12\catcode `\#12\catcode `\^12\catcode `\_12\catcode `\%12\relax}%
\providecommand \@@startlink[1]{}%
\providecommand \@@endlink[0]{}%
\providecommand \url  [0]{\begingroup\@sanitize@url \@url }%
\providecommand \@url [1]{\endgroup\@href {#1}{\urlprefix }}%
\providecommand \urlprefix  [0]{URL }%
\providecommand \Eprint [0]{\href }%
\providecommand \doibase [0]{http://dx.doi.org/}%
\providecommand \selectlanguage [0]{\@gobble}%
\providecommand \bibinfo  [0]{\@secondoftwo}%
\providecommand \bibfield  [0]{\@secondoftwo}%
\providecommand \translation [1]{[#1]}%
\providecommand \BibitemOpen [0]{}%
\providecommand \bibitemStop [0]{}%
\providecommand \bibitemNoStop [0]{.\EOS\space}%
\providecommand \EOS [0]{\spacefactor3000\relax}%
\providecommand \BibitemShut  [1]{\csname bibitem#1\endcsname}%
\let\auto@bib@innerbib\@empty
\bibitem [{\citenamefont {Bringmann}\ and\ \citenamefont
  {Weniger}(2012)}]{Bringmann:2012ez}%
  \BibitemOpen
  \bibfield  {author} {\bibinfo {author} {\bibfnamefont {T.}~\bibnamefont
  {Bringmann}}\ and\ \bibinfo {author} {\bibfnamefont {C.}~\bibnamefont
  {Weniger}},\ }\href {\doibase 10.1016/j.dark.2012.10.005} {\bibfield
  {journal} {\bibinfo  {journal} {Phys.Dark Univ.}\ }\textbf {\bibinfo {volume}
  {1}},\ \bibinfo {pages} {194} (\bibinfo {year} {2012})},\ \Eprint
  {http://arxiv.org/abs/1208.5481} {arXiv:1208.5481 [hep-ph]} \BibitemShut
  {NoStop}%
\bibitem [{\citenamefont {Bergstrom}\ and\ \citenamefont
  {Snellman}(1988)}]{Bergstrom:1988fp}%
  \BibitemOpen
  \bibfield  {author} {\bibinfo {author} {\bibfnamefont {L.}~\bibnamefont
  {Bergstrom}}\ and\ \bibinfo {author} {\bibfnamefont {H.}~\bibnamefont
  {Snellman}},\ }\href {\doibase 10.1103/PhysRevD.37.3737} {\bibfield
  {journal} {\bibinfo  {journal} {Phys.Rev.}\ }\textbf {\bibinfo {volume}
  {D37}},\ \bibinfo {pages} {3737} (\bibinfo {year} {1988})}\BibitemShut
  {NoStop}%
\bibitem [{\citenamefont {Bringmann}\ \emph {et~al.}(2012)\citenamefont
  {Bringmann}, \citenamefont {Huang}, \citenamefont {Ibarra}, \citenamefont
  {Vogl},\ and\ \citenamefont {Weniger}}]{Bringmann:2012}%
  \BibitemOpen
  \bibfield  {author} {\bibinfo {author} {\bibfnamefont {T.}~\bibnamefont
  {Bringmann}}, \bibinfo {author} {\bibfnamefont {X.}~\bibnamefont {Huang}},
  \bibinfo {author} {\bibfnamefont {A.}~\bibnamefont {Ibarra}}, \bibinfo
  {author} {\bibfnamefont {S.}~\bibnamefont {Vogl}}, \ and\ \bibinfo {author}
  {\bibfnamefont {C.}~\bibnamefont {Weniger}},\ }\href {\doibase
  10.1088/1475-7516/2012/07/054} {\bibfield  {journal} {\bibinfo  {journal}
  {JCAP}\ }\textbf {\bibinfo {volume} {1207}},\ \bibinfo {pages} {054}
  (\bibinfo {year} {2012})},\ \Eprint {http://arxiv.org/abs/1203.1312}
  {arXiv:1203.1312 [hep-ph]} \BibitemShut {NoStop}%
\bibitem [{\citenamefont {Weniger}(2012)}]{Weniger:2012}%
  \BibitemOpen
  \bibfield  {author} {\bibinfo {author} {\bibfnamefont {C.}~\bibnamefont
  {Weniger}},\ }\href {\doibase 10.1088/1475-7516/2012/08/007} {\bibfield
  {journal} {\bibinfo  {journal} {JCAP}\ }\textbf {\bibinfo {volume} {1208}},\
  \bibinfo {pages} {007} (\bibinfo {year} {2012})},\ \Eprint
  {http://arxiv.org/abs/1204.2797} {arXiv:1204.2797 [hep-ph]} \BibitemShut
  {NoStop}%
\bibitem [{\citenamefont {Pullen}\ \emph {et~al.}(2007)\citenamefont {Pullen},
  \citenamefont {Chary},\ and\ \citenamefont {Kamionkowski}}]{Pullen:2006sy}%
  \BibitemOpen
  \bibfield  {author} {\bibinfo {author} {\bibfnamefont {A.~R.}\ \bibnamefont
  {Pullen}}, \bibinfo {author} {\bibfnamefont {R.-R.}\ \bibnamefont {Chary}}, \
  and\ \bibinfo {author} {\bibfnamefont {M.}~\bibnamefont {Kamionkowski}},\
  }\href {\doibase 10.1103/PhysRevD.76.063006, 10.1103/PhysRevD.83.029904}
  {\bibfield  {journal} {\bibinfo  {journal} {Phys.Rev.}\ }\textbf {\bibinfo
  {volume} {D76}},\ \bibinfo {pages} {063006} (\bibinfo {year} {2007})},\
  \Eprint {http://arxiv.org/abs/astro-ph/0610295} {arXiv:astro-ph/0610295
  [astro-ph]} \BibitemShut {NoStop}%
\bibitem [{\citenamefont {Abdo}\ \emph {et~al.}(2010)\citenamefont {Abdo},
  \citenamefont {Ackermann}, \citenamefont {Ajello}, \citenamefont {Atwood},
  \citenamefont {Baldini} \emph {et~al.}}]{Abdo:2010nc}%
  \BibitemOpen
  \bibfield  {author} {\bibinfo {author} {\bibfnamefont {A.}~\bibnamefont
  {Abdo}}, \bibinfo {author} {\bibfnamefont {M.}~\bibnamefont {Ackermann}},
  \bibinfo {author} {\bibfnamefont {M.}~\bibnamefont {Ajello}}, \bibinfo
  {author} {\bibfnamefont {W.}~\bibnamefont {Atwood}}, \bibinfo {author}
  {\bibfnamefont {L.}~\bibnamefont {Baldini}},  \emph {et~al.},\ }\href
  {\doibase 10.1103/PhysRevLett.104.091302} {\bibfield  {journal} {\bibinfo
  {journal} {Phys.Rev.Lett.}\ }\textbf {\bibinfo {volume} {104}},\ \bibinfo
  {pages} {091302} (\bibinfo {year} {2010})},\ \Eprint
  {http://arxiv.org/abs/1001.4836} {arXiv:1001.4836 [astro-ph.HE]} \BibitemShut
  {NoStop}%
\bibitem [{\citenamefont {Vertongen}\ and\ \citenamefont
  {Weniger}(2011)}]{Vertongen:2011mu}%
  \BibitemOpen
  \bibfield  {author} {\bibinfo {author} {\bibfnamefont {G.}~\bibnamefont
  {Vertongen}}\ and\ \bibinfo {author} {\bibfnamefont {C.}~\bibnamefont
  {Weniger}},\ }\href {\doibase 10.1088/1475-7516/2011/05/027} {\bibfield
  {journal} {\bibinfo  {journal} {JCAP}\ }\textbf {\bibinfo {volume} {1105}},\
  \bibinfo {pages} {027} (\bibinfo {year} {2011})},\ \Eprint
  {http://arxiv.org/abs/1101.2610} {arXiv:1101.2610 [hep-ph]} \BibitemShut
  {NoStop}%
\bibitem [{\citenamefont {Ackermann}\ \emph {et~al.}(2012)\citenamefont
  {Ackermann} \emph {et~al.}}]{Ackermann:2012qk}%
  \BibitemOpen
  \bibfield  {author} {\bibinfo {author} {\bibfnamefont {M.}~\bibnamefont
  {Ackermann}} \emph {et~al.} (\bibinfo {collaboration} {LAT Collaboration}),\
  }\href {\doibase 10.1103/PhysRevD.86.022002} {\bibfield  {journal} {\bibinfo
  {journal} {Phys.Rev.}\ }\textbf {\bibinfo {volume} {D86}},\ \bibinfo {pages}
  {022002} (\bibinfo {year} {2012})},\ \Eprint {http://arxiv.org/abs/1205.2739}
  {arXiv:1205.2739 [astro-ph.HE]} \BibitemShut {NoStop}%
\bibitem [{\citenamefont {Tempel}\ \emph {et~al.}(2012)\citenamefont {Tempel},
  \citenamefont {Hektor},\ and\ \citenamefont {Raidal}}]{tempel:2012ey}%
  \BibitemOpen
  \bibfield  {author} {\bibinfo {author} {\bibfnamefont {E.}~\bibnamefont
  {Tempel}}, \bibinfo {author} {\bibfnamefont {A.}~\bibnamefont {Hektor}}, \
  and\ \bibinfo {author} {\bibfnamefont {M.}~\bibnamefont {Raidal}},\
  }\href@noop {} {\  (\bibinfo {year} {2012})},\ \Eprint
  {http://arxiv.org/abs/1205.1045} {arXiv:1205.1045 [hep-ph]} \BibitemShut
  {NoStop}%
\bibitem [{\citenamefont {Boyarsky}\ \emph {et~al.}(2012)\citenamefont
  {Boyarsky}, \citenamefont {Malyshev},\ and\ \citenamefont
  {Ruchayskiy}}]{Boyarsky:2012ca}%
  \BibitemOpen
  \bibfield  {author} {\bibinfo {author} {\bibfnamefont {A.}~\bibnamefont
  {Boyarsky}}, \bibinfo {author} {\bibfnamefont {D.}~\bibnamefont {Malyshev}},
  \ and\ \bibinfo {author} {\bibfnamefont {O.}~\bibnamefont {Ruchayskiy}},\
  }\href@noop {} {\  (\bibinfo {year} {2012})},\ \Eprint
  {http://arxiv.org/abs/1205.4700} {arXiv:1205.4700 [astro-ph.HE]} \BibitemShut
  {NoStop}%
\bibitem [{\citenamefont {{Su}}\ and\ \citenamefont
  {{Finkbeiner}}(2012)}]{linepaper}%
  \BibitemOpen
  \bibfield  {author} {\bibinfo {author} {\bibfnamefont {M.}~\bibnamefont
  {{Su}}}\ and\ \bibinfo {author} {\bibfnamefont {D.~P.}\ \bibnamefont
  {{Finkbeiner}}},\ }\href@noop {} {\bibfield  {journal} {\bibinfo  {journal}
  {ArXiv e-prints}\ } (\bibinfo {year} {2012})},\ \Eprint
  {http://arxiv.org/abs/1206.1616} {arXiv:1206.1616 [astro-ph.HE]} \BibitemShut
  {NoStop}%
\bibitem [{\citenamefont {Rajaraman}\ \emph {et~al.}(2012)\citenamefont
  {Rajaraman}, \citenamefont {Tait},\ and\ \citenamefont
  {Whiteson}}]{Rajaraman:2012}%
  \BibitemOpen
  \bibfield  {author} {\bibinfo {author} {\bibfnamefont {A.}~\bibnamefont
  {Rajaraman}}, \bibinfo {author} {\bibfnamefont {T.~M.}\ \bibnamefont {Tait}},
  \ and\ \bibinfo {author} {\bibfnamefont {D.}~\bibnamefont {Whiteson}},\
  }\href {\doibase 10.1088/1475-7516/2012/09/003} {\bibfield  {journal}
  {\bibinfo  {journal} {JCAP}\ }\textbf {\bibinfo {volume} {1209}},\ \bibinfo
  {pages} {003} (\bibinfo {year} {2012})},\ \Eprint
  {http://arxiv.org/abs/1205.4723} {arXiv:1205.4723 [hep-ph]} \BibitemShut
  {NoStop}%
\bibitem [{\citenamefont {Dudas}\ \emph {et~al.}(2012)\citenamefont {Dudas},
  \citenamefont {Mambrini}, \citenamefont {Pokorski},\ and\ \citenamefont
  {Romagnoni}}]{Dudas:2012}%
  \BibitemOpen
  \bibfield  {author} {\bibinfo {author} {\bibfnamefont {E.}~\bibnamefont
  {Dudas}}, \bibinfo {author} {\bibfnamefont {Y.}~\bibnamefont {Mambrini}},
  \bibinfo {author} {\bibfnamefont {S.}~\bibnamefont {Pokorski}}, \ and\
  \bibinfo {author} {\bibfnamefont {A.}~\bibnamefont {Romagnoni}},\ }\href
  {\doibase 10.1007/JHEP10(2012)123} {\bibfield  {journal} {\bibinfo  {journal}
  {JHEP}\ }\textbf {\bibinfo {volume} {1210}},\ \bibinfo {pages} {123}
  (\bibinfo {year} {2012})},\ \Eprint {http://arxiv.org/abs/1205.1520}
  {arXiv:1205.1520 [hep-ph]} \BibitemShut {NoStop}%
\bibitem [{\citenamefont {Choi}\ and\ \citenamefont {Seto}(2012)}]{Choi:2012}%
  \BibitemOpen
  \bibfield  {author} {\bibinfo {author} {\bibfnamefont {K.-Y.}\ \bibnamefont
  {Choi}}\ and\ \bibinfo {author} {\bibfnamefont {O.}~\bibnamefont {Seto}},\
  }\href {\doibase 10.1103/PhysRevD.86.043515, 10.1103/PhysRevD.86.089904}
  {\bibfield  {journal} {\bibinfo  {journal} {Phys.Rev.}\ }\textbf {\bibinfo
  {volume} {D86}},\ \bibinfo {pages} {043515} (\bibinfo {year} {2012})},\
  \Eprint {http://arxiv.org/abs/1205.3276} {arXiv:1205.3276 [hep-ph]}
  \BibitemShut {NoStop}%
\bibitem [{\citenamefont {Kyae}\ and\ \citenamefont {Park}(2013)}]{Kyae:2012}%
  \BibitemOpen
  \bibfield  {author} {\bibinfo {author} {\bibfnamefont {B.}~\bibnamefont
  {Kyae}}\ and\ \bibinfo {author} {\bibfnamefont {J.-C.}\ \bibnamefont
  {Park}},\ }\href {\doibase 10.1016/j.physletb.2012.12.041} {\bibfield
  {journal} {\bibinfo  {journal} {Phys.Lett.}\ }\textbf {\bibinfo {volume}
  {B718}},\ \bibinfo {pages} {1425} (\bibinfo {year} {2013})},\ \Eprint
  {http://arxiv.org/abs/1205.4151} {arXiv:1205.4151 [hep-ph]} \BibitemShut
  {NoStop}%
\bibitem [{\citenamefont {Lee}\ \emph {et~al.}(2012)\citenamefont {Lee},
  \citenamefont {Park},\ and\ \citenamefont {Park}}]{Lee:2012}%
  \BibitemOpen
  \bibfield  {author} {\bibinfo {author} {\bibfnamefont {H.~M.}\ \bibnamefont
  {Lee}}, \bibinfo {author} {\bibfnamefont {M.}~\bibnamefont {Park}}, \ and\
  \bibinfo {author} {\bibfnamefont {W.-I.}\ \bibnamefont {Park}},\ }\href
  {\doibase 10.1103/PhysRevD.86.103502} {\bibfield  {journal} {\bibinfo
  {journal} {Phys.Rev.}\ }\textbf {\bibinfo {volume} {D86}},\ \bibinfo {pages}
  {103502} (\bibinfo {year} {2012})},\ \Eprint {http://arxiv.org/abs/1205.4675}
  {arXiv:1205.4675 [hep-ph]} \BibitemShut {NoStop}%
\bibitem [{\citenamefont {Acharya}\ \emph {et~al.}(2012)\citenamefont
  {Acharya}, \citenamefont {Kane}, \citenamefont {Kumar}, \citenamefont {Lu},\
  and\ \citenamefont {Zheng}}]{Acharya:2012}%
  \BibitemOpen
  \bibfield  {author} {\bibinfo {author} {\bibfnamefont {B.~S.}\ \bibnamefont
  {Acharya}}, \bibinfo {author} {\bibfnamefont {G.}~\bibnamefont {Kane}},
  \bibinfo {author} {\bibfnamefont {P.}~\bibnamefont {Kumar}}, \bibinfo
  {author} {\bibfnamefont {R.}~\bibnamefont {Lu}}, \ and\ \bibinfo {author}
  {\bibfnamefont {B.}~\bibnamefont {Zheng}},\ }\href@noop {} {\bibfield
  {journal} {\bibinfo  {journal} {ArXiv e-prints}\ } (\bibinfo {year}
  {2012})},\ \Eprint {http://arxiv.org/abs/1205.5789} {arXiv:1205.5789
  [hep-ph]} \BibitemShut {NoStop}%
\bibitem [{\citenamefont {Ibarra}\ \emph {et~al.}(2012)\citenamefont {Ibarra},
  \citenamefont {Lopez~Gehler},\ and\ \citenamefont {Pato}}]{Ibarra:2012}%
  \BibitemOpen
  \bibfield  {author} {\bibinfo {author} {\bibfnamefont {A.}~\bibnamefont
  {Ibarra}}, \bibinfo {author} {\bibfnamefont {S.}~\bibnamefont
  {Lopez~Gehler}}, \ and\ \bibinfo {author} {\bibfnamefont {M.}~\bibnamefont
  {Pato}},\ }\href {\doibase 10.1088/1475-7516/2012/07/043} {\bibfield
  {journal} {\bibinfo  {journal} {JCAP}\ }\textbf {\bibinfo {volume} {1207}},\
  \bibinfo {pages} {043} (\bibinfo {year} {2012})},\ \Eprint
  {http://arxiv.org/abs/1205.0007} {arXiv:1205.0007 [hep-ph]} \BibitemShut
  {NoStop}%
\bibitem [{\citenamefont {Buckley}\ and\ \citenamefont
  {Hooper}(2012)}]{Buckley:2012}%
  \BibitemOpen
  \bibfield  {author} {\bibinfo {author} {\bibfnamefont {M.~R.}\ \bibnamefont
  {Buckley}}\ and\ \bibinfo {author} {\bibfnamefont {D.}~\bibnamefont
  {Hooper}},\ }\href {\doibase 10.1103/PhysRevD.86.043524} {\bibfield
  {journal} {\bibinfo  {journal} {Phys.Rev.}\ }\textbf {\bibinfo {volume}
  {D86}},\ \bibinfo {pages} {043524} (\bibinfo {year} {2012})},\ \Eprint
  {http://arxiv.org/abs/1205.6811} {arXiv:1205.6811 [hep-ph]} \BibitemShut
  {NoStop}%
\bibitem [{\citenamefont {Chu}\ \emph {et~al.}(2012)\citenamefont {Chu},
  \citenamefont {Hambye}, \citenamefont {Scarna},\ and\ \citenamefont
  {Tytgat}}]{Chu:2012}%
  \BibitemOpen
  \bibfield  {author} {\bibinfo {author} {\bibfnamefont {X.}~\bibnamefont
  {Chu}}, \bibinfo {author} {\bibfnamefont {T.}~\bibnamefont {Hambye}},
  \bibinfo {author} {\bibfnamefont {T.}~\bibnamefont {Scarna}}, \ and\ \bibinfo
  {author} {\bibfnamefont {M.~H.}\ \bibnamefont {Tytgat}},\ }\href {\doibase
  10.1103/PhysRevD.86.083521} {\bibfield  {journal} {\bibinfo  {journal}
  {Phys.Rev.}\ }\textbf {\bibinfo {volume} {D86}},\ \bibinfo {pages} {083521}
  (\bibinfo {year} {2012})},\ \Eprint {http://arxiv.org/abs/1206.2279}
  {arXiv:1206.2279 [hep-ph]} \BibitemShut {NoStop}%
\bibitem [{\citenamefont {Kang}\ \emph {et~al.}(2012)\citenamefont {Kang},
  \citenamefont {Li}, \citenamefont {Li},\ and\ \citenamefont
  {Liu}}]{Kang:2012}%
  \BibitemOpen
  \bibfield  {author} {\bibinfo {author} {\bibfnamefont {Z.}~\bibnamefont
  {Kang}}, \bibinfo {author} {\bibfnamefont {T.}~\bibnamefont {Li}}, \bibinfo
  {author} {\bibfnamefont {J.}~\bibnamefont {Li}}, \ and\ \bibinfo {author}
  {\bibfnamefont {Y.}~\bibnamefont {Liu}},\ }\href@noop {} {\bibfield
  {journal} {\bibinfo  {journal} {ArXiv e-prints}\ } (\bibinfo {year}
  {2012})},\ \Eprint {http://arxiv.org/abs/1206.2863} {arXiv:1206.2863
  [hep-ph]} \BibitemShut {NoStop}%
\bibitem [{\citenamefont {Buchmuller}\ and\ \citenamefont
  {Garny}(2012)}]{Buchmuller:2012}%
  \BibitemOpen
  \bibfield  {author} {\bibinfo {author} {\bibfnamefont {W.}~\bibnamefont
  {Buchmuller}}\ and\ \bibinfo {author} {\bibfnamefont {M.}~\bibnamefont
  {Garny}},\ }\href {\doibase 10.1088/1475-7516/2012/08/035} {\bibfield
  {journal} {\bibinfo  {journal} {JCAP}\ }\textbf {\bibinfo {volume} {1208}},\
  \bibinfo {pages} {035} (\bibinfo {year} {2012})},\ \Eprint
  {http://arxiv.org/abs/1206.7056} {arXiv:1206.7056 [hep-ph]} \BibitemShut
  {NoStop}%
\bibitem [{\citenamefont {Bergstrom}(2012)}]{Bergstrom:2012b}%
  \BibitemOpen
  \bibfield  {author} {\bibinfo {author} {\bibfnamefont {L.}~\bibnamefont
  {Bergstrom}},\ }\href {\doibase 10.1103/PhysRevD.86.103514} {\bibfield
  {journal} {\bibinfo  {journal} {Phys.Rev.}\ }\textbf {\bibinfo {volume}
  {D86}},\ \bibinfo {pages} {103514} (\bibinfo {year} {2012})},\ \Eprint
  {http://arxiv.org/abs/1208.6082} {arXiv:1208.6082 [hep-ph]} \BibitemShut
  {NoStop}%
\bibitem [{\citenamefont {Heo}\ and\ \citenamefont {Kim}(2013)}]{Heo:2012}%
  \BibitemOpen
  \bibfield  {author} {\bibinfo {author} {\bibfnamefont {J.~H.}\ \bibnamefont
  {Heo}}\ and\ \bibinfo {author} {\bibfnamefont {C.}~\bibnamefont {Kim}},\
  }\href {\doibase 10.1103/PhysRevD.87.013007} {\bibfield  {journal} {\bibinfo
  {journal} {Phys.Rev.}\ }\textbf {\bibinfo {volume} {D87}},\ \bibinfo {pages}
  {013007} (\bibinfo {year} {2013})},\ \Eprint {http://arxiv.org/abs/1207.1341}
  {arXiv:1207.1341 [astro-ph.HE]} \BibitemShut {NoStop}%
\bibitem [{\citenamefont {Park}\ and\ \citenamefont {Park}(2013)}]{Park:2012}%
  \BibitemOpen
  \bibfield  {author} {\bibinfo {author} {\bibfnamefont {J.-C.}\ \bibnamefont
  {Park}}\ and\ \bibinfo {author} {\bibfnamefont {S.~C.}\ \bibnamefont
  {Park}},\ }\href {\doibase 10.1016/j.physletb.2012.12.035} {\bibfield
  {journal} {\bibinfo  {journal} {Phys.Lett.}\ }\textbf {\bibinfo {volume}
  {B718}},\ \bibinfo {pages} {1401} (\bibinfo {year} {2013})},\ \Eprint
  {http://arxiv.org/abs/1207.4981} {arXiv:1207.4981 [hep-ph]} \BibitemShut
  {NoStop}%
\bibitem [{\citenamefont {Tulin}\ \emph {et~al.}(2013)\citenamefont {Tulin},
  \citenamefont {Yu},\ and\ \citenamefont {Zurek}}]{Tulin:2012}%
  \BibitemOpen
  \bibfield  {author} {\bibinfo {author} {\bibfnamefont {S.}~\bibnamefont
  {Tulin}}, \bibinfo {author} {\bibfnamefont {H.-B.}\ \bibnamefont {Yu}}, \
  and\ \bibinfo {author} {\bibfnamefont {K.~M.}\ \bibnamefont {Zurek}},\ }\href
  {\doibase 10.1103/PhysRevD.87.036011} {\bibfield  {journal} {\bibinfo
  {journal} {Phys.Rev.}\ }\textbf {\bibinfo {volume} {D87}},\ \bibinfo {pages}
  {036011} (\bibinfo {year} {2013})},\ \Eprint {http://arxiv.org/abs/1208.0009}
  {arXiv:1208.0009 [hep-ph]} \BibitemShut {NoStop}%
\bibitem [{\citenamefont {Asano}\ \emph {et~al.}(2012)\citenamefont {Asano},
  \citenamefont {Bringmann}, \citenamefont {Sigl},\ and\ \citenamefont
  {Vollmann}}]{Asano:2012zv}%
  \BibitemOpen
  \bibfield  {author} {\bibinfo {author} {\bibfnamefont {M.}~\bibnamefont
  {Asano}}, \bibinfo {author} {\bibfnamefont {T.}~\bibnamefont {Bringmann}},
  \bibinfo {author} {\bibfnamefont {G.}~\bibnamefont {Sigl}}, \ and\ \bibinfo
  {author} {\bibfnamefont {M.}~\bibnamefont {Vollmann}},\ }\href@noop {} {\
  (\bibinfo {year} {2012})},\ \Eprint {http://arxiv.org/abs/1211.6739}
  {arXiv:1211.6739 [hep-ph]} \BibitemShut {NoStop}%
\bibitem [{\citenamefont {Cline}(2012)}]{Cline:2012}%
  \BibitemOpen
  \bibfield  {author} {\bibinfo {author} {\bibfnamefont {J.~M.}\ \bibnamefont
  {Cline}},\ }\href {\doibase 10.1103/PhysRevD.86.015016} {\bibfield  {journal}
  {\bibinfo  {journal} {Phys.Rev.}\ }\textbf {\bibinfo {volume} {D86}},\
  \bibinfo {pages} {015016} (\bibinfo {year} {2012})},\ \Eprint
  {http://arxiv.org/abs/1205.2688} {arXiv:1205.2688 [hep-ph]} \BibitemShut
  {NoStop}%
\bibitem [{\citenamefont {Weiner}\ and\ \citenamefont
  {Yavin}(2012)}]{Weiner:2012}%
  \BibitemOpen
  \bibfield  {author} {\bibinfo {author} {\bibfnamefont {N.}~\bibnamefont
  {Weiner}}\ and\ \bibinfo {author} {\bibfnamefont {I.}~\bibnamefont {Yavin}},\
  }\href {\doibase 10.1103/PhysRevD.86.075021} {\bibfield  {journal} {\bibinfo
  {journal} {Phys.Rev.}\ }\textbf {\bibinfo {volume} {D86}},\ \bibinfo {pages}
  {075021} (\bibinfo {year} {2012})},\ \Eprint {http://arxiv.org/abs/1206.2910}
  {arXiv:1206.2910 [hep-ph]} \BibitemShut {NoStop}%
\bibitem [{\citenamefont {Weiner}\ and\ \citenamefont
  {Yavin}(2013)}]{WeinerYavin:2012b}%
  \BibitemOpen
  \bibfield  {author} {\bibinfo {author} {\bibfnamefont {N.}~\bibnamefont
  {Weiner}}\ and\ \bibinfo {author} {\bibfnamefont {I.}~\bibnamefont {Yavin}},\
  }\href {\doibase 10.1103/PhysRevD.87.023523} {\bibfield  {journal} {\bibinfo
  {journal} {Phys.Rev.}\ }\textbf {\bibinfo {volume} {D87}},\ \bibinfo {pages}
  {023523} (\bibinfo {year} {2013})},\ \Eprint {http://arxiv.org/abs/1209.1093}
  {arXiv:1209.1093 [hep-ph]} \BibitemShut {NoStop}%
\bibitem [{\citenamefont {Fan}\ and\ \citenamefont
  {Reece}(2012)}]{FanReece:2012}%
  \BibitemOpen
  \bibfield  {author} {\bibinfo {author} {\bibfnamefont {J.}~\bibnamefont
  {Fan}}\ and\ \bibinfo {author} {\bibfnamefont {M.}~\bibnamefont {Reece}},\
  }\href@noop {} {\bibfield  {journal} {\bibinfo  {journal} {ArXiv e-prints}\ }
  (\bibinfo {year} {2012})},\ \Eprint {http://arxiv.org/abs/1209.1097}
  {arXiv:1209.1097 [hep-ph]} \BibitemShut {NoStop}%
\bibitem [{\citenamefont {Huang}\ \emph {et~al.}(2012)\citenamefont {Huang},
  \citenamefont {Yuan}, \citenamefont {Yin}, \citenamefont {Bi},\ and\
  \citenamefont {Chen}}]{Huang:2012}%
  \BibitemOpen
  \bibfield  {author} {\bibinfo {author} {\bibfnamefont {X.-Y.}\ \bibnamefont
  {Huang}}, \bibinfo {author} {\bibfnamefont {Q.}~\bibnamefont {Yuan}},
  \bibinfo {author} {\bibfnamefont {P.-F.}\ \bibnamefont {Yin}}, \bibinfo
  {author} {\bibfnamefont {X.-J.}\ \bibnamefont {Bi}}, \ and\ \bibinfo {author}
  {\bibfnamefont {X.-L.}\ \bibnamefont {Chen}},\ }\href {\doibase
  10.1088/1475-7516/2012/11/048} {\bibfield  {journal} {\bibinfo  {journal}
  {JCAP}\ }\textbf {\bibinfo {volume} {1211}},\ \bibinfo {pages} {048}
  (\bibinfo {year} {2012})},\ \Eprint {http://arxiv.org/abs/1208.0267}
  {arXiv:1208.0267 [astro-ph.HE]} \BibitemShut {NoStop}%
\bibitem [{\citenamefont {Whiteson}(2012)}]{Whiteson:2012}%
  \BibitemOpen
  \bibfield  {author} {\bibinfo {author} {\bibfnamefont {D.}~\bibnamefont
  {Whiteson}},\ }\href {\doibase 10.1088/1475-7516/2012/11/008} {\bibfield
  {journal} {\bibinfo  {journal} {JCAP}\ }\textbf {\bibinfo {volume} {1211}},\
  \bibinfo {pages} {008} (\bibinfo {year} {2012})},\ \Eprint
  {http://arxiv.org/abs/1208.3677} {arXiv:1208.3677 [astro-ph.HE]} \BibitemShut
  {NoStop}%
\bibitem [{\citenamefont {Cholis}\ \emph {et~al.}(2012)\citenamefont {Cholis},
  \citenamefont {Tavakoli},\ and\ \citenamefont {Ullio}}]{Cholis:2012}%
  \BibitemOpen
  \bibfield  {author} {\bibinfo {author} {\bibfnamefont {I.}~\bibnamefont
  {Cholis}}, \bibinfo {author} {\bibfnamefont {M.}~\bibnamefont {Tavakoli}}, \
  and\ \bibinfo {author} {\bibfnamefont {P.}~\bibnamefont {Ullio}},\ }\href
  {\doibase 10.1103/PhysRevD.86.083525} {\bibfield  {journal} {\bibinfo
  {journal} {Phys.Rev.}\ }\textbf {\bibinfo {volume} {D86}},\ \bibinfo {pages}
  {083525} (\bibinfo {year} {2012})},\ \Eprint {http://arxiv.org/abs/1207.1468}
  {arXiv:1207.1468 [hep-ph]} \BibitemShut {NoStop}%
\bibitem [{\citenamefont {Rao}\ and\ \citenamefont
  {Whiteson}(2013)}]{Rao:2012fh}%
  \BibitemOpen
  \bibfield  {author} {\bibinfo {author} {\bibfnamefont {K.}~\bibnamefont
  {Rao}}\ and\ \bibinfo {author} {\bibfnamefont {D.}~\bibnamefont {Whiteson}},\
  }\href {\doibase 10.1088/1475-7516/2013/03/035} {\bibfield  {journal}
  {\bibinfo  {journal} {JCAP}\ }\textbf {\bibinfo {volume} {1303}},\ \bibinfo
  {pages} {035} (\bibinfo {year} {2013})},\ \Eprint
  {http://arxiv.org/abs/1210.4934} {arXiv:1210.4934 [astro-ph.HE]} \BibitemShut
  {NoStop}%
\bibitem [{\citenamefont {Whiteson}(2013)}]{Whiteson:2013cs}%
  \BibitemOpen
  \bibfield  {author} {\bibinfo {author} {\bibfnamefont {D.}~\bibnamefont
  {Whiteson}},\ }\href@noop {} {\  (\bibinfo {year} {2013})},\ \Eprint
  {http://arxiv.org/abs/1302.0427} {arXiv:1302.0427 [astro-ph.HE]} \BibitemShut
  {NoStop}%
\bibitem [{\citenamefont {Carlson}\ \emph {et~al.}(2013)\citenamefont
  {Carlson}, \citenamefont {Linden}, \citenamefont {Profumo},\ and\
  \citenamefont {Weniger}}]{Carlson:2013vka}%
  \BibitemOpen
  \bibfield  {author} {\bibinfo {author} {\bibfnamefont {E.}~\bibnamefont
  {Carlson}}, \bibinfo {author} {\bibfnamefont {T.}~\bibnamefont {Linden}},
  \bibinfo {author} {\bibfnamefont {S.}~\bibnamefont {Profumo}}, \ and\
  \bibinfo {author} {\bibfnamefont {C.}~\bibnamefont {Weniger}},\ }\href@noop
  {} {\  (\bibinfo {year} {2013})},\ \Eprint {http://arxiv.org/abs/1304.5524}
  {arXiv:1304.5524 [astro-ph.HE]} \BibitemShut {NoStop}%
\bibitem [{\citenamefont {{Bloom}}\ \emph {et~al.}(2013)\citenamefont
  {{Bloom}}, \citenamefont {{Charles}}, \citenamefont {{Izaguirre}},
  \citenamefont {{Snyder}}, \citenamefont {{Albert}}, \citenamefont {{Winer}},
  \citenamefont {{Yang}},\ and\ \citenamefont
  {{Essig}}}]{bloom_charles_fermi_lat_line}%
  \BibitemOpen
  \bibfield  {author} {\bibinfo {author} {\bibfnamefont {E.}~\bibnamefont
  {{Bloom}}}, \bibinfo {author} {\bibfnamefont {E.}~\bibnamefont {{Charles}}},
  \bibinfo {author} {\bibfnamefont {E.}~\bibnamefont {{Izaguirre}}}, \bibinfo
  {author} {\bibfnamefont {A.}~\bibnamefont {{Snyder}}}, \bibinfo {author}
  {\bibfnamefont {A.}~\bibnamefont {{Albert}}}, \bibinfo {author}
  {\bibfnamefont {B.}~\bibnamefont {{Winer}}}, \bibinfo {author} {\bibfnamefont
  {Z.}~\bibnamefont {{Yang}}}, \ and\ \bibinfo {author} {\bibfnamefont
  {R.}~\bibnamefont {{Essig}}},\ }\href@noop {} {\bibfield  {journal} {\bibinfo
   {journal} {ArXiv e-prints}\ } (\bibinfo {year} {2013})},\ \Eprint
  {http://arxiv.org/abs/1303.2733} {arXiv:1303.2733 [astro-ph.HE]} \BibitemShut
  {NoStop}%
\bibitem [{\citenamefont {Albert}(2012)}]{Albert:Talk}%
  \BibitemOpen
  \bibfield  {author} {\bibinfo {author} {\bibfnamefont {A.}~\bibnamefont
  {Albert}},\ }\href@noop {} {\  (\bibinfo {year} {2012})},\ \bibinfo {note}
  {{P}resentation on Fourth International Fermi Symposium, 2 Nov 2013,
  Monterey}\BibitemShut {NoStop}%
\bibitem [{\citenamefont {Weniger}(2013)}]{Weniger:2013dya}%
  \BibitemOpen
  \bibfield  {author} {\bibinfo {author} {\bibfnamefont {C.}~\bibnamefont
  {Weniger}},\ }\href@noop {} {\  (\bibinfo {year} {2013})},\ \Eprint
  {http://arxiv.org/abs/1303.1798} {arXiv:1303.1798 [astro-ph.HE]} \BibitemShut
  {NoStop}%
\bibitem [{\citenamefont {Abdo}\ \emph {et~al.}(2009)\citenamefont {Abdo} \emph
  {et~al.}}]{FermiLimb}%
  \BibitemOpen
  \bibfield  {author} {\bibinfo {author} {\bibfnamefont {A.}~\bibnamefont
  {Abdo}} \emph {et~al.},\ }\href {\doibase 10.1103/PhysRevD.80.122004}
  {\bibfield  {journal} {\bibinfo  {journal} {Phys.Rev.}\ }\textbf {\bibinfo
  {volume} {D80}},\ \bibinfo {eid} {122004} (\bibinfo {year} {2009})},\ \Eprint
  {http://arxiv.org/abs/0912.1868} {arXiv:0912.1868 [astro-ph.HE]} \BibitemShut
  {NoStop}%
\bibitem [{\citenamefont {Finkbeiner}\ \emph {et~al.}(2013)\citenamefont
  {Finkbeiner}, \citenamefont {Su},\ and\ \citenamefont
  {Weniger}}]{finkbeiner_systematics}%
  \BibitemOpen
  \bibfield  {author} {\bibinfo {author} {\bibfnamefont {D.~P.}\ \bibnamefont
  {Finkbeiner}}, \bibinfo {author} {\bibfnamefont {M.}~\bibnamefont {Su}}, \
  and\ \bibinfo {author} {\bibfnamefont {C.}~\bibnamefont {Weniger}},\ }\href
  {\doibase 10.1088/1475-7516/2013/01/029} {\bibfield  {journal} {\bibinfo
  {journal} {JCAP}\ }\textbf {\bibinfo {volume} {1301}},\ \bibinfo {pages}
  {029} (\bibinfo {year} {2013})},\ \Eprint {http://arxiv.org/abs/1209.4562}
  {arXiv:1209.4562 [astro-ph.HE]} \BibitemShut {NoStop}%
\bibitem [{\citenamefont {Hektor}\ \emph {et~al.}(2012)\citenamefont {Hektor},
  \citenamefont {Raidal},\ and\ \citenamefont {Tempel}}]{Hektor:2012ev}%
  \BibitemOpen
  \bibfield  {author} {\bibinfo {author} {\bibfnamefont {A.}~\bibnamefont
  {Hektor}}, \bibinfo {author} {\bibfnamefont {M.}~\bibnamefont {Raidal}}, \
  and\ \bibinfo {author} {\bibfnamefont {E.}~\bibnamefont {Tempel}},\
  }\href@noop {} {\  (\bibinfo {year} {2012})},\ \Eprint
  {http://arxiv.org/abs/1209.4548} {arXiv:1209.4548 [astro-ph.HE]} \BibitemShut
  {NoStop}%
\bibitem [{1229445()}]{Fermi-LAT:2013jsa}%
  \BibitemOpen
  \bibfield  {author} {1229445,\ }\href@noop {} {\  (\bibinfo {year} {2013})},\
  \Eprint {http://arxiv.org/abs/1304.6082} {arXiv:1304.6082 [astro-ph.HE]}
  \BibitemShut {NoStop}%
\bibitem [{\citenamefont {Gillessen}\ \emph {et~al.}(2011)\citenamefont
  {Gillessen}, \citenamefont {Genzel}, \citenamefont {Fritz}, \citenamefont
  {Quataert}, \citenamefont {Alig} \emph {et~al.}}]{Gillessen:2011aa}%
  \BibitemOpen
  \bibfield  {author} {\bibinfo {author} {\bibfnamefont {S.}~\bibnamefont
  {Gillessen}}, \bibinfo {author} {\bibfnamefont {R.}~\bibnamefont {Genzel}},
  \bibinfo {author} {\bibfnamefont {T.}~\bibnamefont {Fritz}}, \bibinfo
  {author} {\bibfnamefont {E.}~\bibnamefont {Quataert}}, \bibinfo {author}
  {\bibfnamefont {C.}~\bibnamefont {Alig}},  \emph {et~al.},\ }\href@noop {} {\
   (\bibinfo {year} {2011})},\ \Eprint {http://arxiv.org/abs/1112.3264}
  {arXiv:1112.3264 [astro-ph.GA]} \BibitemShut {NoStop}%
\bibitem [{\citenamefont {Phifer}\ \emph {et~al.}(2013)\citenamefont {Phifer},
  \citenamefont {Do}, \citenamefont {Meyer}, \citenamefont {Ghez},
  \citenamefont {Witzel} \emph {et~al.}}]{Phifer:2013taa}%
  \BibitemOpen
  \bibfield  {author} {\bibinfo {author} {\bibfnamefont {K.}~\bibnamefont
  {Phifer}}, \bibinfo {author} {\bibfnamefont {T.}~\bibnamefont {Do}}, \bibinfo
  {author} {\bibfnamefont {L.}~\bibnamefont {Meyer}}, \bibinfo {author}
  {\bibfnamefont {A.}~\bibnamefont {Ghez}}, \bibinfo {author} {\bibfnamefont
  {G.}~\bibnamefont {Witzel}},  \emph {et~al.},\ }\href@noop {} {\  (\bibinfo
  {year} {2013})},\ \Eprint {http://arxiv.org/abs/1304.5280} {arXiv:1304.5280
  [astro-ph.GA]} \BibitemShut {NoStop}%
\bibitem [{\citenamefont {Bergstrom}\ \emph {et~al.}(2012)\citenamefont
  {Bergstrom}, \citenamefont {Bertone}, \citenamefont {Conrad}, \citenamefont
  {Farnier},\ and\ \citenamefont {Weniger}}]{Bergstrom:2012vd}%
  \BibitemOpen
  \bibfield  {author} {\bibinfo {author} {\bibfnamefont {L.}~\bibnamefont
  {Bergstrom}}, \bibinfo {author} {\bibfnamefont {G.}~\bibnamefont {Bertone}},
  \bibinfo {author} {\bibfnamefont {J.}~\bibnamefont {Conrad}}, \bibinfo
  {author} {\bibfnamefont {C.}~\bibnamefont {Farnier}}, \ and\ \bibinfo
  {author} {\bibfnamefont {C.}~\bibnamefont {Weniger}},\ }\href {\doibase
  10.1088/1475-7516/2012/11/025} {\bibfield  {journal} {\bibinfo  {journal}
  {JCAP}\ }\textbf {\bibinfo {volume} {1211}},\ \bibinfo {pages} {025}
  (\bibinfo {year} {2012})},\ \Eprint {http://arxiv.org/abs/1207.6773}
  {arXiv:1207.6773 [hep-ph]} \BibitemShut {NoStop}%
\end{thebibliography}%

\end{document}